\documentclass{aa}

\usepackage{txfonts}
\usepackage{graphicx}
\usepackage{booktabs}
\usepackage{siunitx}
\usepackage{multirow}

\usepackage[colorlinks=true,allcolors=blue]{hyperref}

\newcommand{\ffarcs}{\mbox{\ensuremath{.\!\!^{\prime\prime}}}}

\newcolumntype{L}[1]{>{\raggedright\let\newline\\\arraybackslash\hspace{0pt}}m{#1}}
\newcolumntype{C}[1]{>{\centering\let\newline\\\arraybackslash\hspace{0pt}}m{#1}}

\begin{document}

\title{Polarized scattered light from self-luminous exoplanets}
\subtitle{Three-dimensional scattering radiative transfer with ARTES}

\author{
T.~Stolker\inst{1}
\and M.~Min\inst{2,1}
\and D.~M.~Stam\inst{3}
\and P.~Molli\`{e}re\inst{4}
\and C.~Dominik\inst{1}
\and L.\,B.\,F.\,M.~Waters\inst{2,1}
}

\institute{
Anton Pannekoek Institute for Astronomy, University of Amsterdam, Science Park 904, 1098 XH Amsterdam, The Netherlands\\
\email{T.Stolker@uva.nl}
\and SRON Netherlands Institute for Space Research, Sorbonnelaan 2, 3584 CA Utrecht, The Netherlands
\and Aerospace Engineering Department, Technical University Delft, Kluyverweg 1, 2629 HS, Delft, The Netherlands
\and Max-Planck-Institut f\"{u}r Astronomie, K\"{o}nigstuhl 17, D-69117 Heidelberg, Germany
}

\date{Received ?; accepted ?}

\abstract
{Direct imaging has paved the way for atmospheric characterization of young and self-luminous gas giants. Scattering in a horizontally-inhomogeneous atmosphere causes the disk-integrated polarization of the thermal radiation to be linearly polarized, possibly detectable with the newest generation of high-contrast imaging instruments.}
{We aim to investigate the effect of latitudinal and longitudinal cloud variations, circumplanetary disks, atmospheric oblateness, and cloud particle properties on the integrated degree and direction of polarization in the near-infrared. We want to understand how 3D atmospheric asymmetries affect the polarization signal in order to assess the potential of infrared polarimetry for direct imaging observations of planetary-mass companions.}
{We have developed a three-dimensional Monte Carlo radiative transfer code (ARTES) for scattered light simulations in (exo)planetary atmospheres. The code is applicable to calculations of reflected light and thermal radiation in a spherical grid with a parameterized distribution of gas, clouds, hazes, and circumplanetary material. A gray atmosphere approximation is used for the thermal structure.}
{The disk-integrated degree of polarization of a horizontally-inhomogeneous atmosphere is maximal when the planet is flattened, the optical thickness of the equatorial clouds is large compared to the polar clouds, and the clouds are located at high altitude. For a flattened planet, the integrated polarization can both increase or decrease with respect to a spherical planet which depends on the horizontal distribution and optical thickness of the clouds. The direction of polarization can be either parallel or perpendicular to the projected direction of the rotation axis when clouds are zonally distributed. Rayleigh scattering by submicron-sized cloud particles will maximize the polarimetric signal whereas the integrated degree of polarization is significantly reduced with micron-sized cloud particles as a result of forward scattering. The presence of a cold or hot circumplanetary disk may also produce a detectable degree of polarization ($\lesssim$1\%) even with a uniform cloud layer in the atmosphere.}
{}


\maketitle

\section{Introduction}\label{sec:introduction}

High-contrast imaging observations have enabled the direct detection of young and self-luminous gas giant exoplanets at large orbital radii around nearby stars \citep[e.g.,][]{chauvin2004,marois2008,lagrange2009,rameau2013a}. The thermal radiation from directly imaged exoplanets is spatially resolved from the central star which makes them prime targets for atmospheric characterization through photometry \citep[e.g.,][]{kuzuhara2013,janson2013}, medium- and low-resolution spectroscopy \citep[e.g.,][]{konopacky2013,barman2015,macintosh2015}, and high-resolution spectroscopy combined with adaptive optics \citep{snellen2014}. Scattering by atmospheric gases and particles polarizes the planetary radiation, both the thermal radiation of a planet and the reflected stellar light. The newest generation of high-contrast imaging instruments, including SPHERE \citep[Spectro-Polarimetric High-contrast Exoplanet REsearch;][]{beuzit2008} and GPI \citep[Gemini Planet Imager;][]{macintosh2008}, provide the opportunity to measure the infrared polarization signal from self-luminous gas giant exoplanets.

Polarimetric observations of planetary atmospheres date back to the pioneering work on solar system planets by \citet{lyot1929}. The power of polarization of reflected light as a diagnostic was first demonstrated by \citet{hansen1974a} with their determination of the composition and size of the cloud particles in Venus' atmosphere \citep[see also][]{kattawar1971b}. The potential of this technique to detect and characterize exoplanets in reflected light has been widely recognized and several authors have provided numerical and analytical predictions \citep[e.g.,][]{seager2000,stam2004,buenzli2009,madhusudhan2012,karalidi2013}. A first detection of polarized reflected light from an exoplanet has been claimed for HD~189733b \citep{berdyugina2008,berdyugina2011}, a hot-Jupiter that remains spatially unresolved from its star, but the measurements, which require the planet to have a very high degree of polarization, have not yet been confirmed by others \citep{wiktorowicz2009,wiktorowicz2015,bott2016}. Thermal radiation from a planetary atmosphere can also be polarized when it has been scattered. However, the disk-integrated polarization from the thermal photons will be negligible unless scattering occurs in an atmosphere that deviates from spherical symmetry \citep{sengupta2010,marley2011,dekok2011}. Therefore, a polarization measurement may provide information on the oblateness of an atmosphere or the presence of horizontal cloud variations (e.g., bands or patches).

Linear polarization of ultracool field dwarfs has been measured in multiple surveys. For example, \citet{menard2002} detected polarization degrees up to 0.2\% in the $I$ band with a potential trend of increasing polarization from late-M to mid-L type dwarfs because of the presence of larger amounts of dust in the photospheres of cooler dwarfs. The trend was confirmed by \citet{osorio2005} who measured polarization degrees in the range of 0.2--2.5\% caused by possible cloud inhomogeneities, rotationally-induced oblateness, or in some cases the presence of a dusty disk or envelope. Near-infrared polarimetry by \citet{miles2013} showed that the fast rotating dwarfs from their sample (M7 through T2 spectral types) had on average a larger polarization degree than the moderately rotating dwarfs \citep[see also][]{goldman2009,tata2009,osorio2011,miles2016}. Polarimetric observations of self-luminous gas giants and brown dwarf companions are technically more demanding than observations of field dwarfs because of the companion-to-star flux contrast and the small angular separations involved. Additionally, a companion with a 1\% degree of polarization will be two orders of magnitude fainter in polarized light compared to the total intensity of the companion at the same wavelength. A recent attempt was made by \citet{jensen2016} who placed with GPI at the Gemini South telescope an upper limit of 2.4\% on the degree of polarization of the T5.5 brown dwarf companion in the HD~19467 system.

Several numerical studies have been undertaken to investigate the effect of atmospheric asymmetries on the degree of polarization of self-luminous gas giants and brown dwarfs, showing typical values of linear polarization up to 1--2\% in the optical and near-infrared. \citet{sengupta2010} argued that the $I$-band degree of polarization of field L~dwarfs can be explained by rotationally-induced oblateness and a uniform cloud layer, suggesting low surface gravity \citep[see also][]{sengupta2005}. The same authors have extended their work to gas giant exoplanets and pointed out the potential of infrared polarization measurements of directly imaged planets for the study of their surface gravity and cloud inhomogeneities \citep{marley2011}. The effect of horizontal inhomogeneities in exoplanet atmospheres was studied by \citet{dekok2011}, including the dependence of the degree of polarization on a temperature gradient in the atmosphere. Furthermore, the authors pointed out that the angle of polarization yields the direction of the planet's projected spin axis in case of zonally symmetric cloud structures. More recently, \citet{sengupta2016} showed that an exomoon transit of a self-luminous gas giant may produce a change in polarization up to a few tens of percent in the near-infrared.

In this work, we will investigate the effect of 3D atmospheric variations on the disk-integrated degree and direction of polarization of self-luminous gas giant exoplanets, in continuing line with the initial work by \citet{dekok2011} on horizontal inhomogeneities. We will use an intrinsic three-dimensional (3D) atmospheric radiative transfer code (ARTES), specifically developed for scattering calculations in (exo)planetary atmospheres, to study the effect of horizontal cloud variations, circumplanetary disks, atmospheric oblateness, and cloud particle properties. A gray atmosphere approximation is used for the thermal structure of the atmosphere, and a simplified parametrization for the clouds and circumplanetary disks is implemented. This allows us to directly control the 3D optical depth variations and assess their impact on the polarization signal. The radiative transfer includes a full treatment of multiple scattering and polarization.

In Sect.~\ref{sec:artes}, we provide a basic description of the radiative transfer code and the underlying physics. In Sect.~\ref{sec:self_luminous}, we explore the effect of atmospheric oblateness, horizontally-inhomogeneous cloud structures, circumplanetary disks, and cloud particle properties on the disk-integrated degree and direction of polarization of self-luminous gas giants. In Sect.~\ref{sec:discussion_conclusions}, we discuss the results and the potential of infrared polarimetry for high-contrast exoplanet observations. In Appendix~\ref{sec:benchmarks}, we demonstrate the precision of ARTES with the results of multiple radiative transfer benchmark calculations both of thermal radiation and reflected light. In Appendix~\ref{sec:model_spectra}, we compare the calculated emission spectra with those of a physical atmosphere model. In Appendix~\ref{sec:horizontal}, we investigate the effect of differential transport of horizontally propagating radiation in an example atmosphere with non-uniform clouds.

\section{ARTES: 3D scattering radiative transfer}\label{sec:artes}

\subsection{Code description}\label{sec:code_description}

The Atmospheric Radiative Transfer for Exoplanet Science\footnote{\url{https://github.com/tomasstolker/ARTES}} (ARTES) code applies 3D Monte Carlo radiative transfer to solve wavelength and phase angle-dependent scattering calculations in (exo)planetary atmospheres. The code is written in Fortran~90 and parallelized with OpenMP for use on multi-core processors. The intrinsic 3D setup of ARTES allows for modeling of arbitrary atmospheric structures without the need for any approximations of or assumptions on 3D scattering processes occurring in planetary atmospheres.

The thermal structure of the atmosphere is currently not calculated but a pressure-temperature ($P$-$T$) profile can be provided as input from which the volume mixing ratios and opacities of the gas species are obtained from a pre-calculated grid of opacities (see Sect.~\ref{sec:opacities_matrices}). The gas is assumed to be in hydrostatic equilibrium such that the mass density of the gas is determined by the $P$-$T$ profile. Inhomogeneous cloud and haze layers are included by providing additional opacities to the gas structure of the atmosphere.

Photon packages (hereafter referred to as photons) are emitted from either the stellar photosphere, for reflected light calculations, or from within the atmosphere of the planet itself, for thermal emission calculations. ARTES tracks the emitted photons through a 3D spherical grid and photons are being scattered and absorbed by the gas and cloud particles in the grid cells until a photon is either exiting the atmospheric grid or is being absorbed. The peel-off technique is applied each time a photon scatters in the atmosphere, meaning that the photon location is projected on a detector and the energy is accordingly weighted with the scattering phase function and optical depth \citep{yusef1984}. In regions of high optical depth, the modified random walk approximation is used such that photons can diffuse outward in a small number of computational steps instead of taking an inefficient random walk \citep{min2009}. By using many photons, high signal-to-noise spectra, phase curves, and images can be obtained.

\subsection{Monte Carlo radiative transfer}\label{sec:monte_carlo}

Monte Carlo radiative transfer is a powerful method for calculations in multi-dimensional inhomogeneous environments which has been applied to various astrophysical radiative transfer problems such as protoplanetary disks, supernovae, molecular clouds, and planetary atmospheres \citep[e.g.,][]{goncalves2004,pinte2006,kasen2006,hood2008}. The radiative transfer is computed by stochastically emitting photons and tracing them along their trajectories while sampling various quantities, such as emission directions and scattering angles, from probability distribution functions (PDFs). Sampling of a quantity $x_0$ from a PDF, $p(x)$, is achieved with the normalized cumulative distribution function (CDF), $\psi (x_0)$, which can be calculated by integrating the PDF \citep[e.g.,][]{whitney2011},
\begin{equation}\label{eq:cdf}
\psi (x_0) = \frac{\int_a^{x_0} p(x) dx}{\int_a^b p(x) dx}.
\end{equation}
The inversion of the CDF enables sampling of $x_0$ in the range from $a$ to $b$ with a random number generator. Depending on the complexity of the PDF, this integral can often not be solved analytically in which case ARTES samples from pre-tabulated CDFs which are linearly interpolated.

The precision of a Monte Carlo radiative transfer simulation is determined by the number of photons that is used. The energy of the photons is conserved when radiative equilibrium is enforced by directly reemitting photons that have been absorbed \citep{lucy1999,bjorkman2001}. In that case, the signal-to-noise ratio (S/N) of the measured flux is given by ${\rm S/N} = N_{\rm det}^{1/2}$ (i.e., the Poisson noise of the photons), where $N_{\rm det}$ is the total number of photons that have arrived on a specific detector pixel. This is not the case in ARTES, because energy is peeled from the photon packages as they propagate through the atmospheric grid in order to enhance the S/N. Whenever a thermal photon is emitted or a scattering event occurs, the photon location and energy are projected toward the detector with the energy weighted by the optical depth and phase function. Therefore, the error on the Stokes parameters, $\delta X$, is given by the standard error of the sum, $\delta X = \sigma_{\rm X} \sqrt{N_{\rm det}}$, where $\sigma_{\rm X}$ is the standard deviation of the projected Stokes~$I$, $Q$, $U$, and $V$ photons. The degree of polarization is defined as
\begin{equation}\label{eq:polarization_degree}
P = \frac{\sqrt{Q^2+U^2}}{I},
\end{equation}
where $PI = \sqrt{Q^2+U^2}$ is the polarized intensity. The fractional error on the degree of polarization is derived by propagating the error on the individual Stokes parameters,
\begin{equation}\label{eq:error}
\frac{\delta P}{P} = \left[ \left( \frac{\delta PI}{PI} \right)^2 + \left( \frac{\delta I}{I} \right)^2 \right]^{1/2},
\end{equation}
with the error on the polarized intensity given by
\begin{equation}
\delta PI = \left[ \frac{ \left( Q \delta Q \right)^2 + \left( U \delta U \right)^2}{2\left( Q^2+U^2 \right)} \right]^{1/2}.
\end{equation}
Determining the error on the degree of polarization will be important in Sect.~\ref{sec:infrared_polarization} where we will calculate the disk-integrated polarization signal from self-luminous atmospheres with scattering asymmetries.

\subsection{Photon emission and tracking}\label{sec:photon_emission}

The Monte Carlo photons are packages of equal amount of energy that are either emitted from the stellar photosphere or within the planetary atmosphere itself, with an initial energy of unity. Afterwards, the total amount of energy at the detector plane is scaled to physical units with the energy per photon package,
\begin{equation}
E_\lambda^{\rm photon} = \frac{L_\lambda^{\rm star/planet}}{N_{\rm photon}},
\end{equation}
where $L_\lambda^{\rm star/planet}$ is the monochromatic luminosity of the star or planet, and $N_{\rm photon}$ the total number of emitted photons. For reflected light calculations, ARTES uses the effective temperature, $T_{\rm eff}$, of the star to approximate the emitted stellar flux with a Planck function, $\pi B_\lambda$, at a wavelength $\lambda$. For thermal radiation calculations, the photon energy is determined by the monochromatic luminosity of each grid cell,
\begin{equation}
L_\lambda = 4 \pi \rho_{\rm i} \kappa_\lambda^{\rm abs} B_\lambda (T_{\rm i}) V_{\rm i},
\end{equation}
where $\rho_{\rm i}$ is the mass density of a cell $i$, $\kappa_\lambda^{\rm abs}$ the absorption opacity, $B_\lambda (T_{\rm i})$ the Planck function for cell temperature $T_{\rm i}$, and $V_{\rm i}$ the cell volume. The temperature and gas density in the grid cells is set by the $P$-$T$ profile. The density of the cloud particles is added manually and the particles are assumed to have the same temperature as the gas. The luminosity of each $i$-th cell is weighted by a factor
\begin{equation}
W_{\rm i} = \frac{ \sum\limits_{i=1}^{n} V_{\rm i} \kappa_\lambda^{\rm abs} B_\lambda (T_{\rm i}) }{ V_{\rm i} \kappa_\lambda^{\rm abs} B_\lambda (T_{\rm i}) },
\end{equation}
where $n$ is the total number of grid cells, and the energy of the emitted photons is weighted by the reciprocal of $W_{\rm i}$. In this way, an equal number of photons is emitted from each cell while the total energy of the emitted photons over all cells is conserved.

For reflected light calculations, photons are emitted from the stellar photosphere toward the planetary atmosphere, which is a valid assumption in the case $R_{\rm p} \ll D$, with $R_{\rm p}$ the planet radius and $D$ the orbital radius of the planet. The photons are initially unpolarized because the stellar integrated flux from a quiet main sequence star is unpolarized down to a $10^{-6}$ fractional polarization level \citep{kemp1987}. For thermal calculations, a CDF is constructed from the cell luminosities of the atmospheric grid from which the cell of emission is sampled for each photon. Next, a random location in the grid cell and an isotropic emission direction are sampled as starting point. In order to enhance the number of photons that scatter from high-altitude cloud layers, we use an adjusted probability function for the thermal emission direction \citep{gordon1987},
\begin{equation}
p(\zeta) = \frac{\sqrt{1-\epsilon^2}}{\pi(1+\epsilon\cos\zeta)},
\end{equation}
where $\zeta$ is the local polar angle of emission ($\zeta=0$ corresponds to radially downward and $\zeta=\pi$ to radially upward) and $0 \leq \epsilon < 1$ is an input parameter that sets the asymmetry of the emission direction. A larger value of $\epsilon$ will increase the number of photons that are emitted in radially upward direction with a weighted Stokes vector to ensure an isotropically distributed energy flux.

ARTES keeps track of the location and direction of the photons in the atmospheric grid and determines the distance to a next interaction point. The probability that a photon traverses an optical depth $\tau$ is $e^{-\tau}$, therefore, from Eq.~\ref{eq:cdf} it is straightforward to derive that a random optical depth can be sampled from
\begin{equation}
\tau = -\log{\chi},
\end{equation}
where $\chi$ is a random number between zero and one. The first scattering event is forced with appropriate weighting of the Stokes vector such that photons scatter at least once in low optical depth regions \citep{mattila1970,wood1999}. At this point, a few scenarios are possible which depend on the optical thickness of the atmosphere. First, a photon can cross the outermost grid boundary without any further interactions in which case a next photon is emitted. Second, the photon can cross the innermost boundary of the grid where it is either absorbed or reflected with the probability of either process depending on the specified surface albedo. Third, the photon traverses the sampled optical depth and is being scattered or absorbed by the local opacity source.

For the last scenario, a random number determines the fate of a photon by comparing with the local single scattering albedo,
\begin{equation}
\omega = \frac{\kappa_{\rm scat}}{\kappa_{\rm scat}+\kappa_{\rm abs}},
\end{equation}
where $\kappa_{\rm scat}$ is the scattering opacity and $\kappa_{\rm abs}$ the absorption opacity. In general, absorption of photons can be treated in two different ways, either by removing the complete photon from the simulation or by weighting the Stokes vector of the photon with the single scattering albedo. We use a compromise between these two methods by introducing a parameter $0 \leq \chi \leq 1$ which controls the number of scatterings for an individual photon. The photon absorption probability,
\begin{equation}
f_{\rm stop} = 1 - \omega^\chi,
\end{equation}
is an input parameter which determines the factor $\gamma=\omega/(1-f_{\rm stop})$ by which the Stokes vector is weighted. The limiting cases are $\chi=0$, which corresponds to no absorption and only weighting of the Stokes vector with the single scattering albedo ($\gamma=\omega$), and $\chi=1$, which corresponds to only photon scattering or absorption without weighting the Stokes vector ($\gamma=1$).

The distance toward the nearest cell boundary is calculated whenever a photon enters a new grid cell or when a photon scatters into a new direction. In this way, the next grid cell will be determined and the optical depth toward the cell boundary. In ellipsoidal coordinates, the following set of equations provides the distance, $s$, to a radial, latitudinal, and longitudinal cell boundary:
\begin{subequations}
\begin{equation}
\begin{aligned}
& (a^2n_x^2+b^2n_y^2+c^2n_z^2) s^2 + 2 (a^2xn_x+b^2yn_y+c^2zn_z) s \\
& + (a^2x^2+b^2y^2+c^2z^2-r_{\rm cell}^2)= 0,
\end{aligned}
\end{equation}
\begin{equation}
\begin{aligned}
& (a^2n_x^2 +b^2n_y^2-c^2n_z^2\tan^2{\theta_{\rm cell}}) s^2 + 2 (a^2xn_x+b^2yn_y \\
& -c^2zn_z \tan^2{\theta_{\rm cell}}) s + (a^2x^2+b^2y^2 -c^2z^2\tan^2{\theta_{\rm cell}})= 0,
\end{aligned}
\end{equation}
\begin{equation}
\begin{aligned}
s = \frac{ax\sin{\phi_{\rm cell}} - by\cos{\phi_{\rm cell}}}{bn_y\cos{\phi_{\rm cell}} - an_x\sin{\phi_{\rm cell}}},
\end{aligned}
\end{equation}
\end{subequations}
where $r_{\rm cell}$ is the distance of a radial cell boundary from the grid center, $\theta_{\rm cell}$ the polar angle of a latitudinal cell boundary, $\phi_{\rm cell}$ the azimuthal angle of an longitudinal cell boundary, $(x,y,z)$ the photon location, $(n_x,n_y,n_z)$ the photon direction, and $a$, $b$, and $c$ the fractional scaling of the photon path in $x$, $y$, and $z$ direction, respectively, which mimics flattening or stretching of the atmospheric grid. For a planet with oblateness $f_{\rm oblate}$, the scaling parameters are simplified to $a = b = ( 1-f_{\rm oblate} )^{-1}$ and $c = 1$.

\subsection{Sampling of scattering angles}\label{sec:sampling_scattering_angles}

The scattering phase function provides the angular distribution of photons scattered by a particle at a given wavelength. For example, Rayleigh scattering occurs for very small particles, $2\pi a \ll \lambda$ ($a$ is the particle radius and $\lambda$ is the photon wavelength), which is close to isotropic \citep[e.g.,][]{bohren1983} whereas larger particles, $2\pi a \gtrsim \lambda$, show a forward scattering peak which can be approximated by a Henyey-Greenstein phase function \citep{henyey1941}. Mie theory describes the scattering properties of homogeneous spherical particles and is often also used as an approximation for the scattering properties of non-spherical particles \citep{mie1908}. The calculation depends solely on the complex refractive index (i.e., composition) and size distribution of the particles \citep[e.g.,][]{hansen1974b,derooij1984}.

Each time a photon scatters during an ARTES simulation, a scattering angle, $\Theta$, and azimuthal direction angle, $\Phi$, are sampled from a CDF that depends on the full Stokes vector, $\mathbf{S}$, the scattering phase function, $P_{11}$, as well as the $P_{12}$, $P_{13}$, and $P_{14}$ polarization terms in the $4 \times 4$ scattering matrix, $\mathbf{R}$. Neglecting polarization while sampling scattering angles will introduce errors in the calculation of the disk-integrated reflected planetary flux \citep{stam2005}. The Stokes vector of a photon is given by
\begin{equation}
\mathbf{S} = \pi\left( \begin{array}{c} I \\ Q \\ U \\ V \end{array} \right),
\end{equation}
where $\pi I$ is the total flux, $\pi\sqrt{Q^2+U^2}$ the linear polarized flux, and $\pi V$ the circular polarized flux. The azimuthal direction angle is sampled first because the CDF of the scattering angle depends on the azimuthal angle. The scattering and azimuthal angle, $\Theta$ and $\Phi$, are then used to rotate the propagation direction of a photon, $(n_{\rm x}, n_{\rm y}, n_{\rm z})$, and to calculate the new values for the polarization parameters of the Stokes vector (see Sect.~\ref{sec:photon_scattering}). The circular polarization component of the Stokes vector can be non-zero due to multiple scattering (depending on the scattering matrix), for example in the case of Mie scattering particles. Although the full Stokes vector is calculated by ARTES, including the circular polarization component, we focus in this study on linear polarization and ignore the circular polarization results.

\subsection{Photon scattering}\label{sec:photon_scattering}

Scattering of photons occurs through a sequence of matrix multiplications \citep[e.g.,][]{chandrasekhar1960,hovenier2004},
\begin{equation}
\mathbf{S} = \mathbf{L}(\pi - i_2) \mathbf{R}(\Theta) \mathbf{L}(-i_1) \mathbf{S'},
\end{equation}
where $\mathbf{S}$ is the Stokes vector after scattering, $\mathbf{S'}$ the Stokes vector before scattering, $\mathbf{L}$ the rotation matrix, and $\mathbf{R}$ the scattering matrix. The rotation matrix rotates the Stokes vector by angles $i_1$ and $i_2$, from the local meridian reference plane toward the local scattering plane and vice versa, respectively, and is given by
\begin{equation}
\mathbf{L}=
\begin{pmatrix}
1 & 0 & 0 & 0 \\
0 & \cos{2\beta} & \sin{2\beta} & 0 \\
0 & -\sin{2\beta} & \cos{2\beta} & 0 \\
0 & 0 & 0 & 1
\end{pmatrix},
\end{equation}
where $\beta$ is the rotation angle between the reference planes for which by definition a positive value corresponds to counter-clockwise rotation when looking along the photon propagation direction. The rotation matrix alters the reference plane for the linear polarization component of the Stokes vector but does not affect the total intensity, Stokes~$I$, the polarized intensity, $PI$, nor the circular polarized intensity, $V$.

The initial rotation angle, $i_1$, is equal to the randomly-sampled azimuthal direction angle, $\Phi$, which rotates the Stokes vector from the local meridian plane toward the scattering plane (see Sect.~\ref{sec:sampling_scattering_angles}). backward rotation to the new meridian plane is determined by the second rotation angle, $i_2$, which is calculated with the spherical law of cosines. The backward rotation is the clockwise or counter-clockwise direction when $i_2$ is smaller or larger than $\pi$, respectively \citep{hovenier2004}.

\subsection{Opacities and scattering matrices}\label{sec:opacities_matrices}

Absorption opacities of molecular and atomic gas have typically a strong dependence on the atmospheric temperature and pressure \citep[e.g.,][]{freedman2008}. ARTES uses a pre-tabulated grid of volume mixing ratios and opacities ($\lambda/\Delta\lambda=100$) for pressures ranging from 1~$\mu$bar up to 300~bar and temperatures in the range of 75--4000~K. The absorption cross sections of the gas species are calculated with the SRON (Netherlands Institute for Space Research) Planetary Atmosphere Retrieval Code (SPARC; Min et al., in prep.) which applies the correlated-$k$ method \citep[e.g.,][]{qiang1992} for a range of gas molecules and atoms retrieved from the HITRAN \citep{rothman2013}, HITEMP \citep{rothman2010}, and ExoMol \citep{tennyson2012,yurchenko2014} line lists. SPARC uses a stochastic sampling approach for the pressure and thermal broadening of the spectral lines. The pressure-broadened line wings are approximated with a Voigt profile and the thermal-broadened wings with a Gaussian profile. The absorption cross sections are averaged in each spectral bin. The equilibrium chemistry model from \citet{molliere2017} has been used to determine the mixing ratios for the same pressure and temperature grid by minimizing the Gibbs free energy. We have neglected non-equilibrium gas chemistry which can also be important in the atmospheres of self-luminous gas giants \citep[e.g.,][]{zahnle2014}.

Scattering of gas molecules and atoms occurs in the Rayleigh limit, therefore, the scattering cross section of the gas increases toward shorter wavelengths with an approximate $\lambda^{-4}$ wavelength dependence \citep[e.g.,][]{liou1980},
\begin{equation}\label{eq:rayleigh}
\sigma_{\rm scat} = \frac{24\pi^3}{N^2 \lambda^4} \left( \frac{n^2-1}{n^2+2} \right)^2 \frac{6+3\delta}{6-7\delta},
\end{equation}
where $n$ is the real part of the refractive index, $N$ the number of molecules per unit volume, and $\delta$ the depolarization factor which accounts for the spatial anisotropy of a particle. The total mass of a gas giant atmosphere is dominated by molecular hydrogen (H$_2$) for which the wavelength-dependent refractive index is given by \citep{cox2000}
\begin{equation}
n_{\rm H_2} = 1.358 \times 10^{-4} \left[ 1+7.52\times10^{-3} \left( \frac{\lambda}{1~\mu{\rm m}} \right)^{-2} \right] + 1.
\end{equation}

Cloud condensates usually form in the troposphere or in the lower part of the stratosphere when the partial vapor pressure of a gas exceeds the saturation vapor pressure. Liquid or solid cloud particles have typical radii of $\sim$1~$\mu$m in the upper part of the troposphere but will grow up to $\sim$100~$\mu$m by sedimentation toward the lower part of the troposphere \citep[e.g.,][]{ackerman2001}. The detailed distribution of the clouds, as well as the particle sizes, is determined by 3D hydrodynamical motions and kinetic cloud formation processes which can produce submicron-sized particles at pressure levels above $\sim$1~bar \citep[e.g.,][]{lee2016}. Haze particles are usually small in size (submicron) and form through photochemistry or other non-equilibrium processes in the stratosphere \citep[e.g.,][]{marley2013}. Here, we do not incorporate detailed cloud formation processes, instead, we use a parameterized distribution of the cloud layers with an empirical size distribution of the particles.

Mie theory can be used to determine the scattering and absorption opacities of spherical and homogeneous cloud particles, as well as the $4 \times 4$ scattering matrix which contains the scattering angle dependent characteristics \citep[e.g.,][]{hansen1974b},
\begin{equation}\label{eq:mie_matrix}
\mathbf{R}=
\begin{pmatrix}
P_{11} & P_{12} & 0 & 0 \\
P_{12} & P_{22} & 0 & 0 \\
0 & 0 & P_{33} & P_{34} \\
0 & 0 & -P_{34} & P_{44}
\end{pmatrix},
\end{equation}
with the phase function given by $P_{11}$ and the single scattering polarization by $-P_{12}/P_{11}$. Eight of the matrix elements are set to zero due to symmetry arguments. The scattering and absorption properties depend on the complex refractive index which contains a real part, $n$, and an imaginary part, $k$, which are the phase velocity ratio and the attenuation coefficient, respectively. The size distribution of the cloud particles is approximated with a gamma distribution \citep{hansen1971},
\begin{equation}\label{eq:gamma_distribution}
n(r) = C r^{(1-3v_{\rm eff})/v_{\rm eff}} e^{-r/(r_{\rm eff}v_{\rm eff})},
\end{equation}
where $C$ is a normalization constant, $r$ the particle radius, $r_{\rm eff}$ the effective radius of the size distribution, and $v_{\rm eff}$ the effective variance (dimensionless). Not all particles are well approximated by Mie theory, in particular their polarization characteristics. For example, a liquid cloud droplet is better approximated as a spherical homogeneous particle than a hazy aggregate. A uniform distribution of hollow spheres (DHS) can be utilized to determine the optical properties for an ensemble of randomly oriented, irregularly shaped particles \citep{min2005}. In that case, the scattering and absorption properties not only depend on the complex refractive index and the size distribution, but also on the maximum volume void fraction, $f_{\rm DHS}$, of the hollow sphere distribution which controls the irregularity ($f_{\rm DHS}=0$ corresponds to homogeneous spheres). The scattering matrix of the DHS has the same symmetries as the scattering matrix in Mie theory (see Eq.~\ref{eq:mie_matrix}).

\section{Self-luminous gas giant exoplanets}\label{sec:self_luminous}

\begin{figure*}
\resizebox{\hsize}{!}{\includegraphics{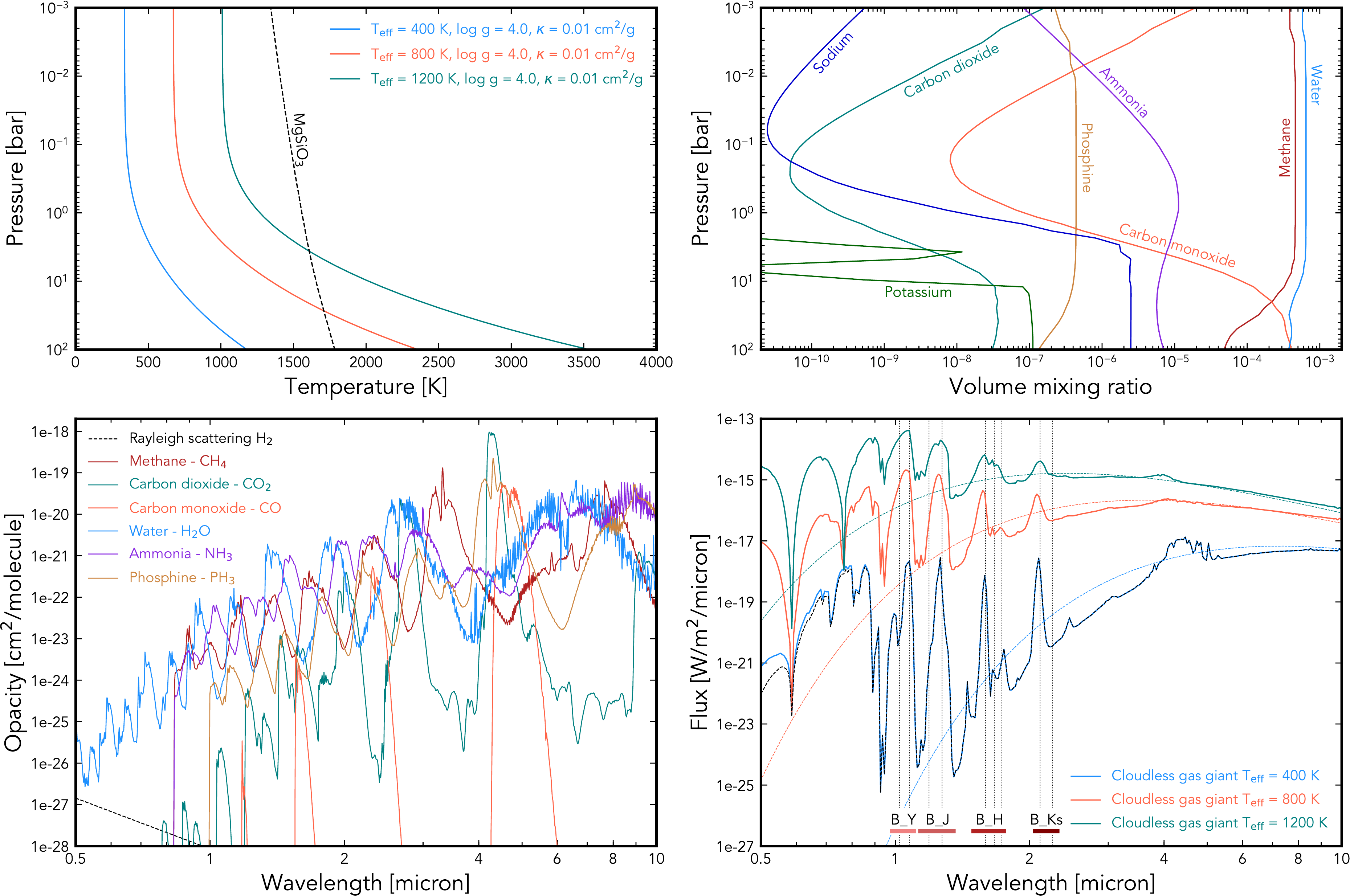}}
\caption{\emph{Top left:} Gray atmosphere pressure-temperature profiles for $T_{\rm eff}=\{400,800,1200\}$~K (solid color lines, see Eq.~\ref{eq:gray_atmosphere}). The condensation curve of enstatite (black dashed line) is shown for solar metallicity \citep{burrows1997}. \emph{Top right:} Mixing ratios of the most abundant gas molecules and atoms for the $T_{\rm eff}=800$~K model \citep{molliere2017}. \emph{Bottom left:} Absorption cross sections of the most abundant gas molecules (solid color lines), calculated for $T=1000$~K and $P=1$~bar, as well as the Rayleigh scattering cross section of H$_2$ (black dashed line). \emph{Bottom right:} Emission spectra of cloudless gas giant atmospheres at a distance of 10~pc with effective temperatures of 400, 800, and 1200~K (solid color lines) together with the Planck functions for the same temperatures (dashed color lines). A comparison spectrum for which scattering is neglected is shown for the 400~K atmosphere (black dashed line). The width of the SPHERE/IRDIS broadband filters are displayed on the bottom of the figure. Vertically dotted lines correspond to the central wavelengths of the IRDIS dual-band filters (from left to right: $Y2$, $Y3$, $J2$, $J3$, $H2$, $H3$, $H4$, $K1$, and $K2$) which have a typical width of 50~nm in the $Y$, $J$, and $H$ bands, and 100~nm in the $K$ band.\label{fig:self_luminous}}
\end{figure*}

Young and self-luminous exoplanets have been detected with high-contrast differential imaging techniques around approximately a dozen of stars. The polarimetric signal from those planets will be non-zero if (i) thermal radiation is scattered and (ii) the polarized intensity is distributed asymmetrically across the planetary disk. In this section, we will use ARTES to investigate the effect of oblateness, horizontally-inhomogeneous clouds, circumplanetary disks, and particle scattering properties on the degree and direction of polarization from self-luminous gas giants. We use a parameterized setup in order to directly control the atmospheric, circumplanetary, and particle properties.

\subsection{Cloudless atmospheres}\label{sec:cloudless}

We start with a description of a cloudless atmosphere model which is the basis for the models with scattering asymmetries in  Sect.~\ref{sec:infrared_polarization}. A simplified thermal structure is used, given by the gray atmosphere approximation \citep[e.g.,][]{hansen2008,guillot2010},
\begin{equation}\label{eq:gray_atmosphere}
T(P) = \left[ \frac{3}{4}T_{\rm eff}^4 \left( \frac{2}{3} + \frac{\kappa_{\rm IR} P}{g} \right) \right]^{1/4},
\end{equation}
where $T_{\rm eff}$ is the effective temperature, $\kappa_{\rm IR}$ the infrared opacity, $P$ the gas pressure, and $g$ the surface gravity. The thermal structure is calculated with Eq.~\ref{eq:gray_atmosphere} for 50~logarithmically spaced pressure layers from 1~mbar down to 100~bar. A constant surface gravity of $\log{g}=4.0$~(cm~s$^{-2}$) is used, which is a typical value inferred from direct imaging observations \citep[e.g.,][]{madhusudhan2011}, and a gray opacity of 0.01~cm$^2$ g$^{-1}$ \citep[e.g.,][]{sharp2007}. We consider a self-luminous exoplanet on a wide orbit such that the contribution of reflected starlight is negligible in the near-infrared. Three example $P$-$T$ profiles are shown in the top left plot of Fig.~\ref{fig:self_luminous} which are calculated for $T_{\rm eff}=\{400,800,1200\}$~K. The upper part of the $P$-$T$ profiles ($\gtrsim 0.1$~bar) is close to isothermal whereas the deeper atmospheric layers are approximately adiabatic.

For each atmospheric layer, we calculated the pressure and temperature-dependent mixing ratios and absorption cross sections of the gas molecules and atoms by linearly interpolating a pre-tabulated grid (see Sect.~\ref{sec:opacities_matrices}). We assumed a solar metallicity atmosphere, ${\rm [M/H]} = 0.0$, and solar carbon-to-oxygen ratio, ${\rm C/O}=0.54$. The top right plot of Fig.~\ref{fig:self_luminous} shows the mixing ratios of the eight most abundant molecules (CH$_4$, CO$_2$, CO, H$_2$O, NH$_3$, PH$_3$) and atoms (Na, K) for a cloudless atmosphere with $T_{\rm eff}=1000$~K. In the cooler parts of the atmosphere ($\lesssim 10$~bar), H$_2$O and CH$_4$ have the largest mixing ratios, whereas CO and H$_2$O are the dominant gas species in the higher temperature layers below $\sim$10~bar. The absorption cross sections are calculated in the wavelength range of 0.5--30~$\mu$m with a spectral resolution of $\lambda/\Delta\lambda=100$. The bottom left plot of Fig.~\ref{fig:self_luminous} shows an example of the opacities of the eight most abundant gas molecules for $P=1$~bar and $T=1000$~K, together with the (pressure and temperature independent) Rayleigh scattering opacity of H$_2$ (see Eq.~\ref{eq:rayleigh}).

The $P$-$T$ profile is used by ARTES to calculate the vertical density structure of the gas (assuming hydrostatic equilibrium) which together with the absorption cross sections and mixing ratios determines the absorption opacity in each atmospheric layer. For the scattering optical thickness, we only consider H$_2$ molecules which are the dominant scattering source in a gas giant atmosphere. Scattering mainly contributes to the spectrum in the wavelength regime below $\sim$1.0~$\mu$m and decreases steeply in the near-infrared. The relative contribution of scattering depends on the atmospheric temperature due to the temperature-dependent absorption component in the single scattering albedo of the gas (see bottom left plot in Fig.~\ref{fig:self_luminous}).

The calculated emission spectra are shown in the bottom right plot of Fig.~\ref{fig:self_luminous} for $T_{\rm eff}=\{400,800,1200\}$~K (see Appendix~\ref{sec:benchmark_emission_spectrum} for a benchmark emission spectrum). The pressure-broadened sodium and potassium resonance doublets are visible at 0.59~$\mu$m and 0.77~$\mu$m, respectively, with increasing strength toward higher temperatures. The largest contribution to the molecular absorption bands in the near-infrared comes from water, with increasing depth of the water band with decreasing temperature. Also methane is important in the considered temperature regime. Rayleigh scattering contributes to the emergent flux only at optical wavelengths. An additional spectrum is calculated in which scattering has been excluded which shows a lower continuum flux (below 0.8~$\mu$m) compared to the 400~K spectrum that fully includes multiple scattering, whereas the difference in the absorption bands is minor. In Appendix~\ref{sec:model_spectra}, we show a comparison of the calculated emission spectra with those from the AMES-Cond atmospheric models by \citet{allard2001}.

\subsection{Infrared polarization from scattering asymmetries}\label{sec:infrared_polarization}

\begin{figure*}
\resizebox{\hsize}{!}{\includegraphics{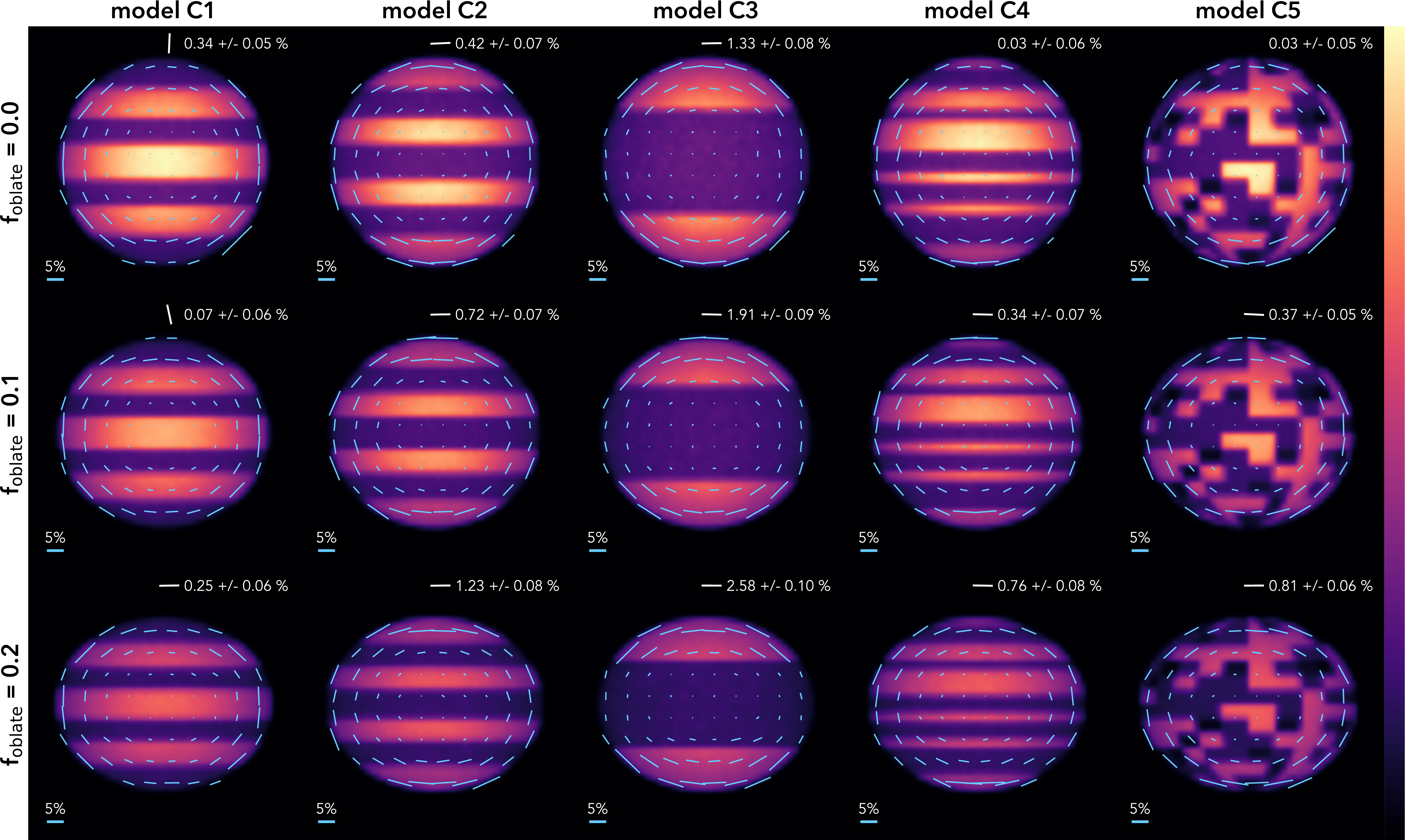}}
\caption{Grid of ARTES models with non-uniform cloud layers. The color of the images shows the total intensity across the planetary disk and the vectors denote the local direction and degree of polarization. The five columns correspond from left to right to ARTES models C1--C5, and the three rows show from top to bottom increasing oblateness, $f_{\rm oblate}=\{0.0,0.1,0.2\}$. All polarization vectors are identically normalized with the length of a 5\% polarization vector shown in the bottom left of each image. The disk-integrated degree of polarization and the $5\sigma$ Monte Carlo uncertainty is shown for each model in the top right of the image. The white vector denotes the direction of polarization of the integrated signal (the vector has been left out in model C4 and C5 with $f_{\rm oblate}=0.0$ because the S/N of the integrated degree of polarization is low). The color scale is identical for all images with the maximum value given by the peak intensity across all models.\label{fig:clouds}}
\end{figure*}

Thermal radiation becomes linearly polarized due to scattering by gas, cloud, or haze particles in a planetary atmosphere. Photometric observations of self-luminous exoplanets measure the emitted flux integrated over the planetary disk. For a spherically shaped planet with plane-symmetric cloud structures, the net Stokes~$Q$ and $U$ flux is zero because the positive and negative lobes across the planetary disk cancel each other. The net polarized intensity will be non-zero when the thermally emitted light is scattered in an atmosphere with an asymmetric distribution of scatterers. Therefore, measuring infrared polarization from a self-luminous exoplanet yields information on the asymmetry of the atmosphere. In this section, we investigate the effect of oblateness, horizontal cloud variations, circumplanetary disks, and the scattering properties of the cloud particles on the disk-integrated degree and direction of polarization of self-luminous gas giants.

\subsubsection{Non-uniform cloud variations}\label{sec:clouds}

We have constructed five ARTES planet models (C1--C5) with a parametric implementation of location and optical thickness of the clouds. We aim to provide a proof of concept of 3D scattering processes that result in a non-zero polarization signal and we leave a more realistic calculation, coupled to a physical atmospheric model for future work. Here, the optical depth variations through the clouds, as well as the scattering properties of the cloud particles are fixed, whereas in Sect.~\ref{sec:single_scattering}, we use a fixed cloud structure but different scattering properties to study the effect of the cloud particles on the integrated polarization signal.

For the thermal structure, we use a $P$-$T$ profile with $T_{\rm eff}=800$~K (see Sect.~\ref{sec:cloudless}) from which the vertical density structure of the gas is calculated, as well as the gas opacities and mixing ratios. The mean molecular weight and surface gravity, which determine the density structure, are set to $\mu=2.3$ (solar) and \mbox{$\log{g} = 4.0$}, respectively, both constant throughout the atmosphere. The models are complemented with additional cloud opacities at specific pressure levels, latitudes, and longitudes. Each model is run for oblateness values of $f_{\rm oblate}=1-\frac{R_{\rm p}}{R_{\rm e}}=\{0.0,0.1,0.2\}$, where $R_{\rm p}$ is the polar radius and $R_{\rm e}$ the equatorial radius of the planet (Saturn has an oblateness of 0.1). The rotationally-induced oblateness of a planet is given by the Darwin-Radau relationship \citep{barnes2003},
\begin{equation}\label{eq:oblateness}
f_{\rm oblate} = \frac{\Omega^2 R_{\rm e}}{g} \left[ \frac{5}{2} \left( 1 - \frac{3I}{2MR_{\rm e}^2} \right)^2 + \frac{2}{5} \right]^{-1}
\end{equation}
where $\Omega$ is the rotation rate in rad~s$^{-1}$, $g$ the surface gravity, $I$ the moment of inertia around the rotational axis, and $M$ the planet mass. As an example, a planet with a rotational period of 8~hr \citep{snellen2014}, a surface gravity of $\log{g}=4.0$, and a mass of 5~$M_{\rm Jup}$ yields an oblateness of 0.05 when the moment of inertia is approximated by a solid sphere. However, in this study we parameterize the oblateness without making an assumption about the planet mass or radius, and rotation rate. For the flattened planets, the vertical structure is scaled from the poles toward the equatorial latitudes. The radiative transfer is monochromatic for which we choose the central wavelength of the SPHERE/IRDIS $H2$ continuum filter, 1.59~$\mu$m (see Fig.~\ref{fig:self_luminous}). The detector plane is placed in the equatorial plane of the planet ($\theta_{\rm det} = 0\degr$) such that the flattening is fully visible.

The scattering properties of the cloud particles are kept constant throughout the grid of $5 \times 3$ models. Submicron-sized cloud particles are used with a size distribution given by Eq.~\ref{eq:gamma_distribution} for which the effective radius and variance are set to 0.1~$\mu$m and 0.05, respectively. We assumed an enstatite (MgSiO$_3$) composition with the complex refractive index in the $H$ band obtained from \citet{dorschner1995}. Enstatite dust grains can form at atmospheric altitudes above $\sim$30~bar in the $T_{\rm eff} = 800$~K atmosphere (see condensation curve Fig.~\ref{fig:self_luminous}). The particles will scatter light in the Rayleigh regime because their size is smaller then the $H$-band wavelength (i.e., $2\pi a \ll \lambda$). Therefore, the phase function is close to isotropic,
\begin{equation}\label{eq:phase_rayleigh}
P_{11}(\cos{\Theta}) = \frac{3}{8} \left( 1+\cos^2{\Theta} \right),
\end{equation}
and the single scattering polarization is a perfectly bell-shaped function with 100\% polarization efficiency at a scattering angle of $90\degr$,
\begin{equation}\label{eq:pol_rayleigh}
-\frac{P_{12}(\cos{\Theta})}{P_{11}(\cos{\Theta})} = -\frac{\cos^2{\Theta}-1}{\cos^2{\Theta}+1}.
\end{equation}
We use Mie theory to compute the opacities and scattering matrices therefore assuming homogeneous, spherical particles. In the $H$ band, the single scattering albedo is 0.995 for the chosen size distribution and composition.

The cloud structures are all located at high altitude ($P_{\rm cloud}=10$~mbar) in a single atmospheric layer but the vertical optical depth through the clouds contains horizontal variations with optical depth values of $\tau_1=1$ and $\tau_2=5$. Additional cloud characteristics of the models are the following (see Fig.~\ref{fig:clouds}):
\begin{itemize}
\item Model C1 contains seven latitudinal cloud regions with cloud optical depth variations that are symmetric with respect to the equator plane. Polar latitudes contain thicker clouds than the equatorial latitudes.
\item Model C2 is similar to model C1, but the cloud optical depth variations have been interchanged such that clouds are thicker around the equator and thin around the poles.
\item Model C3 is a more extreme case with thick clouds only between latitudes of $-30\degr$ and $30\degr$, and thinner clouds in the regions north of $30\degr$ and south of $-30\degr$.
\item Model C4 contains eleven bands of clouds, with variation in width, which are asymmetrically distributed across all latitudes.
\item Model C5 contains an atmosphere with patchy clouds. The atmospheric grid has been divided into 12 latitudinal regions between $-90\degr$ and $90\degr$, and 12 longitudinal regions between $0\degr$ and $180\degr$. We sampled 100 random grid cells at a 10~mbar pressure level and added a $\tau_2=5$ cloud layer in each sampled cell. Cells were allowed to be sampled multiple times. In addition, a uniform cloud layer of $\tau_1=1$ is added across the entire atmosphere at 10~mbar to make sure that no cloudless areas are present.
\end{itemize}

Because of the positive temperature gradient, most energy is emitted from deep in the atmosphere causing a net upward flux. Consequently, the spatially resolved degree of polarization is maximal along the limb of the atmosphere. The polarization degree will be large for high altitude clouds in which case photons have a larger probability of being scattered toward the observer compared to clouds that are located deeper in the atmosphere. In the latter scenario, photons have a larger probability of being absorbed by the surrounding gas, which has a single scattering albedo close to zero in the near-infrared, therefore decreasing the degree of polarization. A net upward flux of photons means that the scattering angle along the limb toward the observer is $\Theta=90\degr$ while the scattering angle, as well as the degree of polarization, decreases toward the center of the planetary disk where photons are scattered in forward direction \citep[see][for a more detailed elaboration on the effect of temperature gradients]{dekok2011}.

Spatially resolved polarization maps of the ARTES models C1--C5 are displayed in Fig.~\ref{fig:clouds} for both spherical and oblate atmospheres. The images show the Stokes~$I$ surface brightness from the planetary disks. The presence of thick clouds results in a smaller flux since lower temperature regions are probed. The disk-integrated polarization levels are provided in the top right of each image together with the 5$\sigma$ Monte Carlo uncertainties (calculated with Eq.~\ref{eq:error}) which are in the range of 0.05--0.10\%. The polarization angle, $\chi$, with respect to the reference plane is defined as
\begin{equation}
\chi = \frac{1}{2}\arctan{\frac{U}{Q}}.
\end{equation}
Across the planetary disks, the high S/N polarization vectors are oriented in azimuthal direction as expected for positively polarizing particles in a spherical or ellipsoidal geometry. The disk-integrated polarization direction is oriented in horizontal direction for all models except in model C1 with $f_{\rm oblate} = 0.0$ and $f_{\rm oblate} = 0.1$ in which case the polarization is vertically oriented (see Fig.~\ref{fig:clouds}). Consequently, for an atmosphere with zonal clouds, the measured direction of polarization can be parallel or perpendicular to the direction of the planet's projected rotation axis.

The degree and direction of polarization of the integrated signal depends on the oblateness, the latitudinal and longitudinal placement of the clouds, as well as their optical thickness (a fixed parameter in the presented models). The presence of thick clouds around the equator (e.g., model C2 and C3) results in a disk-integrated flux that is horizontally polarized because the polarized flux is largest at the polar regions (see Fig.~\ref{fig:clouds}). In addition, a non-zero oblateness will increase the horizontally polarized flux because the polar limb is stretched whereas the equatorial limb decreases in spatial scale. For the spherical atmospheres, the integrated polarization signal is either close to zero because approximately an equal amount of horizontally and vertically polarized flux is obscured by clouds (models C4 and C5), or the polarized flux is non-zero in which case the orientation of the polarization direction is determined by the thickness of the clouds at the equatorial and polar latitudes (vertical in model C1 and horizontal in model C2 and C3).

The combined effect of oblateness and cloud thickness is clearly visible in model C1. For a spherical atmosphere, the integrated polarized flux mainly originates from the equatorial latitudes where the polarization is vertical. Increasing the oblateness will increase the polarized flux from the polar latitudes (even though thick clouds are present) and decrease the flux from the equatorial latitudes. Consequently, the integrated degree of polarization reduces (model C1, $f_{\rm oblate} = 0.1$) until the direction of polarization changes by $90\degr$ after which the integrated degree of polarization will start to rise again (model C1, $f_{\rm oblate} = 0.2$). The maximum amount of polarization occurs with the combined presence of a flattened atmosphere, which enhances the horizontally polarized flux, and equatorial clouds, which reduces the vertically polarized flux. As a result, the degree of polarization is maximal in model C3, ranging from 1.33\% for the spherical atmosphere to 2.58\% for a strongly flattened atmosphere.

All models in Fig.~\ref{fig:clouds} contain high-altitude clouds ($P_{\rm cloud} = 10$~mbar) which causes a relatively large polarized flux because the gas above the cloud deck is optically thin. For comparison, we ran model C3 with the clouds located deeper in the atmosphere, $P_{\rm cloud} = 100$~mbar, and all other parameters the same. In that case, the disk-integrated degree of polarization is $1.26\pm0.02\%$, $1.74\pm0.02\%$, and $2.30\pm0.02\%$ for an planet oblateness of 0.0, 0.1, and 0.2, respectively. Lowering the clouds to higher pressure levels will weaken the effect because a larger fraction of unpolarized light is emitted from above the cloud deck which reduces of the degree of polarization.

\subsubsection{The effect of a circumplanetary disk}\label{sec:circumplanetary_disk}

\begin{figure*}
\resizebox{\hsize}{!}{\includegraphics{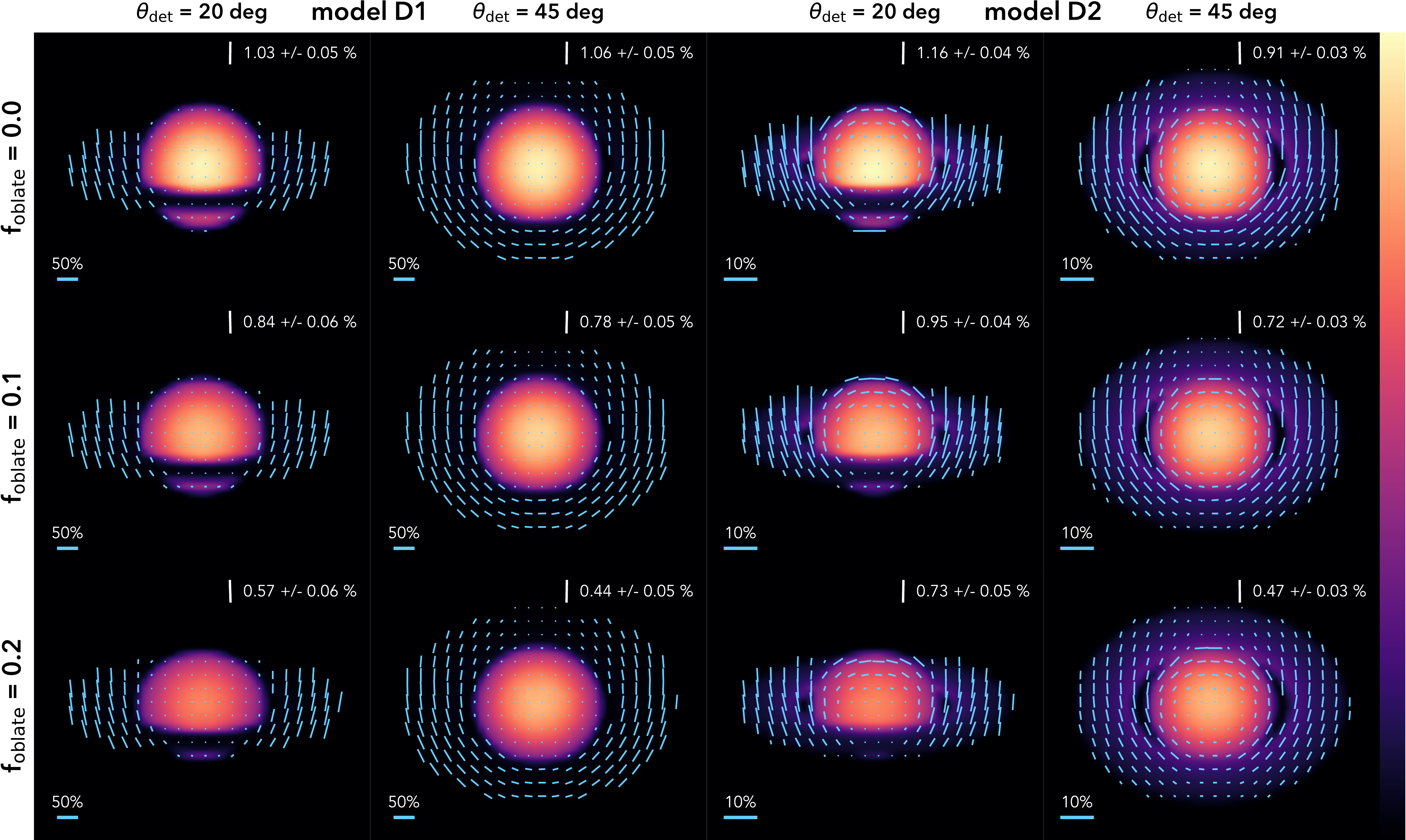}}
\caption{Grid of ARTES infrared polarization models with a circumplanetary disk. Models D1 and D2 contain a cold and hot circumplanetary disk, respectively. Two different viewing direction are shown, $\theta_{\rm det}=20\degr$ and $\theta_{\rm det}=45\degr$. See the caption of Fig.~\ref{fig:clouds} for additional details.\label{fig:circumplanetary}}
\end{figure*}

Another scenario in which the net infrared polarization from a self-luminous gas giant will be non-zero is with the presence of a circumplanetary disk. We have constructed two models (D1 and D2) that contain circumplanetary material in the equatorial plane of the planet. The inner radius of the circumplanetary disk is located at $2 \times 10^4$~km above the atmosphere and extends to $1 \times 10^5$~km (comparable in size to Saturn's ring system). The thermal structure of the atmosphere is the same as before but the cloud layer at 10~mbar does not contain any horizontal variations and has an optical thickness of unity. The cloud particle properties are identical to the non-uniform models in Sect.~\ref{sec:clouds}, that is, submicron-sized enstatite grains. The detector is placed with an offset of $\theta_{\rm det} = 20\degr$ or $\theta_{\rm det} = 45\degr$ from the equatorial plane such that part of the planetary disk is obscured by the circumplanetary disk.

We distinguish between between two different scenarios: (i) a circumplanetary disk (model D1) which is cold ($T_{\rm disk}=100$~K), has a vertical optical depth of $\tau_{\rm in} = 2.0$ and $\tau_{\rm out} = 3.5$ at the inner and outer edge, respectively, has an opening angle of $0.2\degr$, and contains submicron-sized enstatite grains and no gas, and (ii) a circumplanetary disk (model D2) which is hot $T_{\rm disk} = 750$~K, optically thick ($\tau_{\rm in} = 20$ and $\tau_{\rm out} = 34$), has an opening angle of $5\degr$, contains submicron-sized enstatite grains with a dust-to-gas ratio of 0.1, and contains gas with a constant absorption opacity of 0.01~cm$^2$ g$^{-1}$. In model D1, the thermal emission is dominated by the planet photosphere and the non-zero polarization signal is determined by the partial obscuration of the planetary disk and the scattering of atmospheric photons from the circumplanetary disk. Additionally in model D2, the gas in the circumplanetary disk emits a significant amount of radiation which becomes polarized through scattering by dust grains both in the disk and the planet atmosphere. The radiative transfer is computed at 2.11~$\mu$m (SPHERE/IRDIS $K1$ continuum filter; see Fig.~\ref{fig:self_luminous}).

Figure~\ref{fig:circumplanetary} displays the Stokes~$I$ images with corresponding polarization maps for the grid of $2 \times 3 \times 2$ models. Two different viewing angles are used and the oblateness values are the same as in Fig.~\ref{fig:clouds}. In model D1, the disk-integrated polarization is $\sim$1\% when the atmosphere is spherical. For the highly inclined disk ($\theta_{\rm det}=20\degr$), scattering of atmospheric photons from the circumplanetary disk increases the vertically polarized flux but part of the vertically polarized flux from the atmosphere gets obscured by the disk which counteracts the effect. For the mildly inclined disk ($\theta_{\rm det}=20\degr$) around the spherical atmosphere, the net vertically polarized flux is smaller but the disk obscures also a small fraction of the south pole which decreases the horizontally polarized flux. The integrated degree of polarization is largest for the spherical atmospheres and decreases with increasing oblateness. Even though the obscuration of the south pole increases with oblateness when $\theta_{\rm det}=20\degr$, the horizontally polarized flux increases at the north pole with respect to the spherical atmosphere which causes a decrease of the integrated degree of polarization with increasing oblateness. The decrease is steeper for the low inclination disks because a smaller fraction of the south pole gets obscured.

The integrated polarization signal of model D2 is determined by the combined effect of the obscuration of the planetary disk, scattering of thermal photons in the circumplanetary disk, and scattering of atmospheric photons from the disk and vice versa. The optical depth between the circumplanetary disk surface and midplane is large, consequently, photons will be randomly polarized when scattered from deep in the disk toward the detector. The contribution of disk photons to the local polarized disk flux is therefore determined by photons that originate from close to the disk surface of which the net polarization direction is in vertical direction (see model D2 in Fig.~\ref{fig:circumplanetary}). Multiple scattering causes the degree of polarization of the scattered disk photons to be relatively low, in contrast to the atmospheric photons that scatter with a high degree of polarization from the circumplanetary disk. A detector at a larger latitude (i.e., toward the poles, $\theta_{\rm det}=45\degr$) will image the circumplanetary disk with a stronger circular symmetry such that the local polarized flux is smaller compared to a more edge-on viewing direction ($\theta_{\rm det}=20\degr$). As a result, the contribution of scattered disk photons to the polarized flux decreases while the contribution of atmospheric photons scattering from the circumplanetary disk increases (similar to model D1). For a pole-on viewing angle, the spatially resolved polarized flux caused by scattering of thermal disk photons is zero because of the local scattering symmetry, except at the edges of the disk. The integrated polarized flux decreases with increasing oblateness because of the increase of horizontally polarized flux from the visible pole (i.e., similar to model D1).

We used a simplified circumplanetary disk structure with a single temperature and density, therefore, neglecting any vertical and radial gradients and we assumed an homogeneous distribution of submicron-sized grains throughout the disk. Although the scattered light signal is mainly determined by the grains in the uppermost layer of the circumplanetary disk where grains are expected to be small, larger grains might be stratified closed to the disk midplane. Also, the radius of a circumplanetary accretion disk can be a significant fraction of a Hill radius \citep[$\sim$0.4~$R_{\rm H}$;][]{martin2011} in which case the atmospheric polarized flux might be negligible depending on the temperature of both the planet photosphere and the circumplanetary disk, as well as the observed wavelength.

\subsubsection{Dependence on scattering properties}\label{sec:single_scattering}

\begin{table*}
\caption{Infrared polarization: dependence on scattering properties}
\label{table:single_scattering}
\centering
\bgroup
\def\arraystretch{1.35}
\begin{tabular}{C{1cm} | L{7.8cm} | C{2.4cm} C{2.4cm} C{2.4cm}}
\hline\hline
\multirow{2}{*}{Model} & \multirow{2}{*}{Cloud particle} & \multicolumn{3}{c}{Degree of polarization\tablefootmark{e}} \\
\cline{3-5}
& & $f_{\rm oblate} = 0.0$ & $f_{\rm oblate} = 0.1$ & $f_{\rm oblate} = 0.2$ \\
\hline
S1 & Mie, MgSiO$_3$\tablefootmark{a}, $r_{\rm eff}=0.1$~$\mu$m, $v_{\rm eff}=0.05$ & 0.42 +/- 0.07 \% & 0.72 +/- 0.07 \% & 1.23 +/- 0.08 \% \\
S2 & Mie, MgSiO$_3$, $r_{\rm eff}=0.1$~$\mu$m, $v_{\rm eff}=0.05$, $\lambda = 1.67$~$\mu$m\tablefootmark{b} & 0.21 +/- 0.04 \% & 0.38 +/- 0.04 \% & 0.59 +/- 0.04 \% \\
S3 & Henyey-Greenstein, $g_{\rm HG}=0.5$, $\omega_{\rm cloud}=1$ & 0.15 +/- 0.04 \% & 0.26 +/- 0.05 \% & 0.47 +/- 0.05 \% \\
S4 & Henyey-Greenstein, $g_{\rm HG}=0.9$, $\omega_{\rm cloud}=1$ & 0.06 +/- 0.08 \% & 0.08 +/- 0.09 \% & 0.19 +/- 0.10 \% \\
S5 & Mie, MgSiO$_3$, $r_{\rm eff}=1.0$~$\mu$m, $v_{\rm eff}=0.1$ & 0.07 +/- 0.03 \% & 0.07 +/- 0.04 \% & 0.16 +/- 0.04 \% \\
S6 & DHS\tablefootmark{c}, MgSiO$_3$, $r_{\rm eff}=1.0$~$\mu$m, $v_{\rm eff}=0.1$, $f_{\rm DHS}=0.8$\tablefootmark{d} & 0.06 +/- 0.03 \% & 0.14 +/- 0.04 \% & 0.21 +/- 0.04 \% \\
\bottomrule
\end{tabular}
\egroup
\tablefoot{\\
\tablefoottext{a}{Complex refractive indices obtained from \citet{dorschner1995}.}\\
\tablefoottext{b}{Central wavelength of the SPHERE/IRDIS $H3$ filter which is sensitive to methane absorption. For the other models, we used the central wavelength of the SPHERE/IRDIS $H2$ continuum filter, $\lambda = 1.59$~$\mu$m.}\\
\tablefoottext{c}{Irregularly shaped particle properties are approximated with a distribution on hollow spheres \citep[DHS;][]{min2005}.}\\
\tablefoottext{d}{The maximum volume void fraction for the DHS.}\\
\tablefoottext{e}{Integrated degree of polarization (see Eq.~\ref{eq:polarization_degree}) and 5$\sigma$ Monte Carlo uncertainty (see Eq.~\ref{eq:error}).}\\
}
\end{table*}

\begin{figure}
\resizebox{\hsize}{!}{\includegraphics{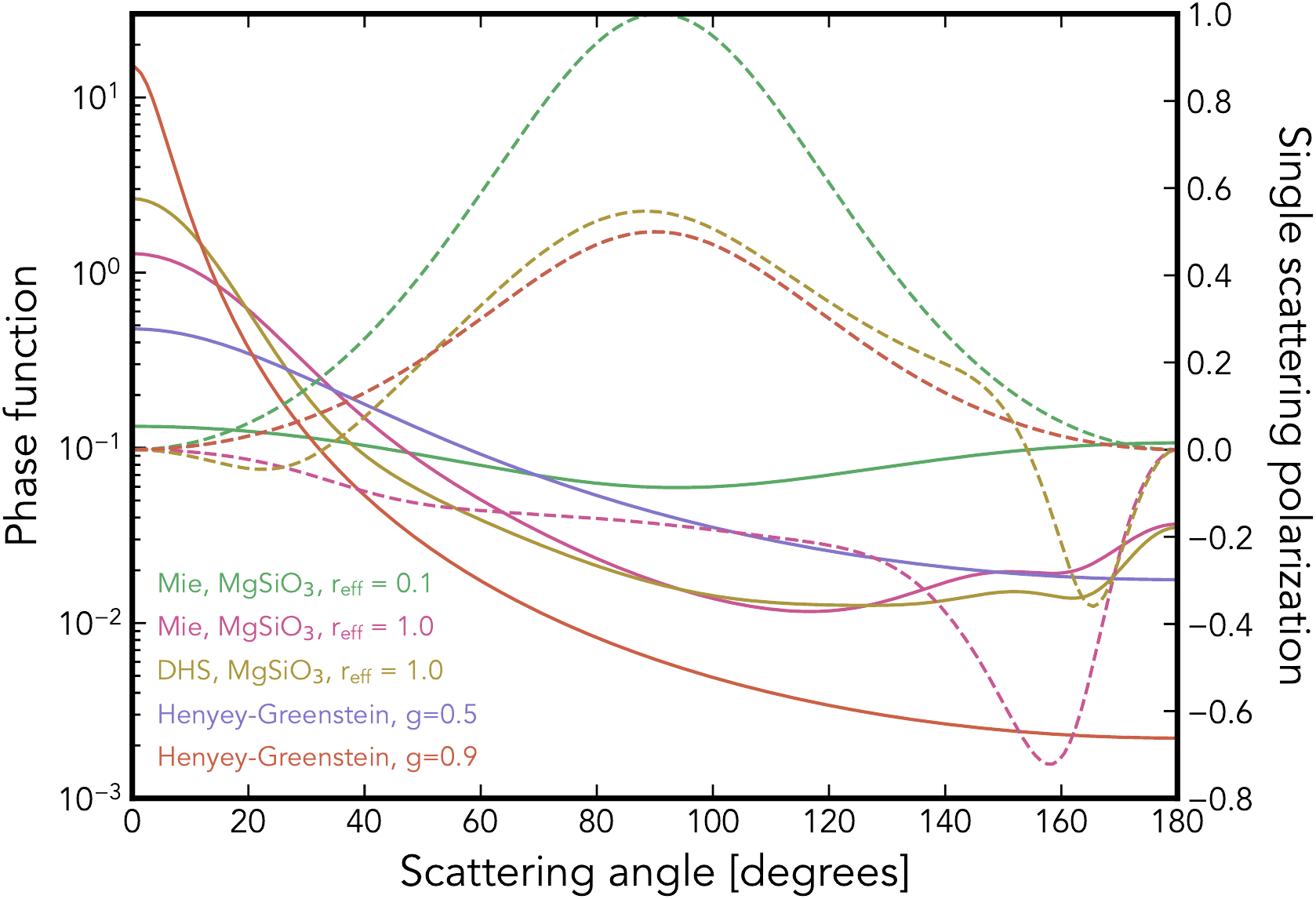}}
\caption{Phase functions (solid lines) and single scattering polarization (dashed lines) of the cloud particles used to study the dependence on the scattering properties. The models and particle properties are listed in Table~\ref{table:single_scattering}. The single scattering polarization is identically parameterized for the two different Henyey-Greenstein asymmetry parameters.\label{fig:single_scattering}}
\end{figure}

In Sects.~\ref{sec:clouds} and \ref{sec:circumplanetary_disk}, we fixed the scattering properties of the cloud particles in order to assess the dependence of the polarimetric signal on spatial variations of the cloud structures and the presence of circumplanetary disks. Here, we fix the atmospheric structure and construct six models (S1--S6) with a variation of particle properties to investigate the dependence of the polarimetric signal on the chosen scattering properties. We use the atmospheric structure of model C2 which contains a distribution of thin clouds ($\tau_1=1$) and thick clouds ($\tau=5$) at a pressure level of 10~mbar. All models are monochromatic and calculated in the $H$-band continuum except for model S2 which is calculated in a methane absorption band. The main characteristics of the cloud particles are the following (see also Table~\ref{table:single_scattering}):

\begin{itemize}
\item Model S1 is identical to C2 and consists of submicron-sized enstatite particles, therefore, the phase function is approximately isotropic (see Eq.~\ref{eq:phase_rayleigh}) and the single scattering polarization is a perfect bell-shape function (see Eq.~\ref{eq:pol_rayleigh}) which maximizes the degree of polarization. The radiative transfer is computed at a continuum wavelength, $\lambda=1.59$~$\mu$m (SPHERE/IRDIS $H2$ filter).
\item Model S2 is identical to model S1, but the radiative transfer is computed within a methane absorption band, $\lambda=1.67$~$\mu$m (SPHERE/IRDIS $H3$ filter).
\item Model S3 contains particles with a Henyey-Greenstein parametrization of the phase function \citep{henyey1941},
\begin{equation}\label{eq:henyey_greenstein}
P_{11}(\cos{\Theta}) = \frac{1-g^2}{(1+g^2-2g\cos{\Theta})^{3/2}},
\end{equation}
where $g_{\rm HG}$ is the scattering asymmetry parameter which is set to 0.5. The single scattering polarization is parameterized by a bell-shaped curve similar to Eq.~\ref{eq:pol_rayleigh} but with the peak normalized to 50\% \citep{white1979}. The single scattering albedo is set to unity, $\omega_{\rm cloud}=1$.
\item Model S4 is similar to S3 but contains cloud particles with a stronger forward scattering phase function with the asymmetry parameter is set to $g_{\rm HG}=0.9$.
\item Model S5 is similar to model S1 but contains micron-sized instead of submicron-sized particles. Their size distribution is described by Eq.~\ref{eq:gamma_distribution} with the effective radius and variance set to 1.0~$\mu$m and 0.1, respectively. Opacities and scattering matrices are calculated with Mie theory.
\item Model S6 is similar to S5 but a distribution of hollow spheres \citep[DHS;][]{min2005} is used to approximate the opacities and scattering matrices of irregularly shaped particles with the maximum volume void fraction, $f_{\rm DHS}$, set to 0.8 \citep{woitke2016}.
\end{itemize}
Figure~\ref{fig:single_scattering} displays a comparison of the phase functions and single scattering polarization curves of the cloud particles.

All models are again calculated for a spherically shaped atmosphere and two values of non-zero oblateness. The results of the $6 \times 3$ model grid are presented in Table~\ref{table:single_scattering}. As expected, the maximum amount of polarization is obtained for the submicron-sized cloud particles which scatter in the Rayleigh regime (model S1). The integrated degree of polarization is considerably lower in model S2 for which the larger gas opacities cause the photosphere to be located at higher altitudes, therefore, closer to the cloud deck. This means that a larger fraction of the flux is unpolarized because it is emitted above the cloud deck and the cloud deck is less asymmetrically irradiated compared to model S2 for which most of the flux originated from below the cloud deck.

For the models with Henyey-Greenstein particles, the polarization is lower because of both the smaller peak polarization and the asymmetry in the phase function which causes a larger fraction of the photons to be scattered upward in radial direction for which the single scattering polarization is small. As a result, the disk-integrated polarization decreases with increasing asymmetry parameter. The phase function in the $H$ band for micron-sized Mie scattering particles, as well as a DHS with similar-sized particles, contains also a forward scattering peak (see Fig.~\ref{fig:single_scattering}). The single scattering polarization on the other hand shows strong differences between DHS and Mie theory. For a DHS, the single scattering polarization is approximately bell-shaped with a fractional polarization peak of $\sim$0.5, whereas the polarization is overall negative for spherical particles (Mie theory). The polarization vector of the disk-integrated signal has an horizontal orientation (similar to those in Fig.~\ref{fig:clouds}) in all models except model S5. In model S5, the integrated polarization vector is oriented in vertical direction because of the difference in sign of the single scattering polarization (see Fig.~\ref{fig:single_scattering}).

\section{Discussion and conclusions}\label{sec:discussion_conclusions}

In this work, we presented a new 3D Monte Carlo radiative transfer code, ARTES, that can be used for wavelength and phase angle-dependent scattering calculations in (exo)planetary atmospheres, both of reflected light and thermal radiation. The code has been carefully benchmarked with the results from several other radiative transfer codes (see Appendix~\ref{sec:benchmarks}). Multiple scattering and polarization are fully implemented. We have computed spatially resolved polarization maps and the corresponding disk-integrated polarization signal from self-luminous gas giants with a parameterized atmospheric structure to investigate the effect of horizontally-inhomogeneous clouds (see Sect.~\ref{sec:clouds}) and the presence of a circumplanetary disk (see Sect.~\ref{sec:circumplanetary_disk}). In addition, we have studied the effect of the scattering properties of the cloud particles (see Sect.~\ref{sec:single_scattering}).

\subsection{Rotation, atmospheric dynamics, clouds, and circumplanetary material}\label{sec:dynamics_clouds}

Thermal radiation from exoplanets, brown dwarfs, and stars becomes linearly polarized by scattering processes occurring in their atmosphere \citep[e.g.,][]{sengupta2001,khan2006}. For a spatially-resolved planetary disk as shown in Fig.~\ref{fig:clouds}, the polarization is maximal around the limb where the scattering angles toward the observer are $90\degr$ \citep[see also][]{dekok2011}. The disk-integrated signal will be polarized if scattering occurs asymmetrically across the planetary disk. Therefore, infrared polarimetry can be used to constrain the rotationally-induced oblateness of a planet or brown dwarf \citep{sengupta2010,marley2011}, as well as horizontal inhomogeneities, the presence of a circumplanetary disk, and the projected direction of the planet's rotational axis in case of zonally symmetric clouds \citep{dekok2011}. The main requirement is the altitude of the clouds which has to be high enough to imprint a detectable polarization signature.

The oblateness of a planet depends on its surface gravity, rotational period, and the internal distribution of mass. Therefore, combined constraints of polarimetry and spectroscopy allow for a better understanding on the internal structure of a planet \citep{marley2011}. Rotation also affects the dynamics in the atmosphere which can result in strong winds and an inhomogeneous distribution of clouds, for example zonal or patchy cloud structures such as those in the atmospheres of Jupiter and Saturn. Cloud condensates in the atmospheres of young and self-luminous planets are composed of refractory materials such as silicates and iron \citep[see e.g.,][]{marley2013}. In Sect.~\ref{sec:clouds}, we determined that stratospheric clouds and hazes may cause polarization levels up to 1--2\%, although strongly dependent on the oblateness, cloud optical depth, scattering properties, and wavelength. The effect of tropospheric clouds ($P_{\rm cloud} \lesssim 100$~mbar) on the disk-integrated polarization is smaller and might be challenging to detect. Three-dimensional global circulation models of self-luminous gas giants predict atmospheric winds both in vertical and horizontal direction that are driven by the rotation of the planet \citep{showman2013} which may circulate (sub)micron-sized cloud particles to millibar pressure levels \citep{lee2016}.

At high altitudes, cloud or haze particles are likely small and possibly aggregate-like which will scatter light with a high degree of polarization because the single scattering polarization is determined by the size of the constituents and not by the aggregate size \citep{west1991,min2016}. Observational evidence for submicron-sized dust grains is provided by the unusual red colors of some L dwarfs which can be explained by a layer of silicate haze in the upper parts of their atmosphere \citep{yang2015,hiranaka2016}. The refractive index affects both the single scattering albedo and polarization efficiency. For example, the single scattering albedo is close to unity for silicates and the single scattering polarization is approximately bell-shaped \citep[e.g.,][]{volten2001}, whereas carbon-rich material has a stronger absorbing efficiency and a polarization curve that is overall lower and deviating from being bell-shaped \citep{munoz2006}. In this study, we used a simplified, parameterized thermal structure and implementation of the clouds, whereas the formation or cloud condensates and photochemical hazes is a complex process which is controlled by many aspects of the atmosphere such as the thermal structure, atmospheric dynamics, and (non-equilibrium) chemistry \citep[see e.g.,][]{helling2008}.

All gas and ice giants in our solar system possess circumplanetary rings, with Saturn clearly having the densest ring system. Therefore, we may also expect similar ring systems around exoplanets, with an extreme case being the 0.6~au ring system proposed by \citet{kenworthy2015} as an explanation for the peculiar photometry of J1407. For a planet young enough ($\lesssim 200$~Myr) to be detected in the near-infrared, a cold circumplanetary disk or ring system can produce a degree of polarization of 0.5--1.0\% even if the planet atmosphere is spherical and uniformly distributed by clouds (see model D1 in Sect.~\ref{sec:circumplanetary_disk}). Additionally, scattering of thermal photons in a hot circumplanetary disk will also contribute to the integrated polarization signal depending on the viewing geometry and observed wavelength (see model D2 in Sec.~\ref{sec:circumplanetary_disk}). A few embedded protoplanets have been directly detected, possibly surrounded by a circumplanetary accretion disk \citep{quanz2015,sallum2015}. High-contrast infrared polarimetry of forming protoplanets will be challenging because those planets are still embedded in or surrounded by a circumstellar disk, therefore, the scattered stellar light might locally dominate over the planetary signal. This technique will likely have a larger potential for more evolved companions that orbit in a dust depleted circumstellar environment, yet, show evidence of circumplanetary material (see Sect.\ref{sec:potential_targets}).

\subsection{Direct polarimetric imaging of companions: opportunities and challenges}

Directly imaged exoplanets are a sparse population of wide orbit gas giants that have been detected with high-contrast adaptive optics (AO) instruments \citep[for a review, see][]{bowler2016}. Although, assuming hot-start evolutionary models, the occurrence rate of 5--13~$M_{\rm Jup}$ planets at orbital distances of 30--300~au is only $0.6^{+0.7}_{-0.5}$\% \citep{bowler2016}, directly imaged exoplanets are key targets for atmospheric characterization through photometry and integral field spectroscopy \citep[e.g.,][]{morzinski2015,skemer2016}, as well as AO-assisted high-resolution spectroscopy \citep{snellen2014}. High-contrast infrared polarimetry has been recognized by several authors as a potentially valuable technique for characterization of exoplanets \citep{marley2011,dekok2011}, yet, the measurement is challenging because of the required sensitivity and absolute polarimetric accuracy to measure polarization levels below 1\%. Consequently, no companions have been detected in polarized light until today.

An opportunity is brought forward with the installment of the spectro-polarimetric imaging instruments GPI \citep{macintosh2008} and SPHERE \citep{beuzit2008} which may provide the required polarimetric precision to detect gas giant and brown dwarf companions in polarized infrared light \citep[e.g.,][]{wiktorowicz2014}. The polarimetric imaging mode of GPI is implemented with a Wollaston prism beamsplitter that can replace the spectral dispersing prism in the integral field spectrograph, and an additional rotating half-wave plate \citep{perrin2010,perrin2015}. Recently, \citet{jensen2016} determined with GPI an 2.4\% upper limit on the degree of polarization of the HD~19467B companion (T5.5~brown dwarf) at a separation of 1\ffarcs65 and $\Delta H = 12.45$~mag contrast. Also $\beta$~Pic was observed with GPI in polarimetric imaging mode which revealed with high S/N $\beta$~Pic~b in Stokes~$I$ but the planet was not recovered in polarized intensity after subtracting a model disk image from the data given that the planet is located close to the major axis of the disk \citep{millar2015}.

Polarimetric imaging in the near-infrared with SPHERE is provided by the IRDIS differential imaging camera. A beam splitter and set of polarizers separates the incoming beam into two beams with orthogonal polarization directions \citep{langlois2014}. Dual-polarimetric imaging (DPI) with IRDIS is offered in field stabilized mode which keeps the direction of polarization fixed on the detector. The contrast of IRDIS DPI observations is mainly limited by the differential aberrations between the two IRDIS channels and the correction of the instrumental polarization which is introduced upstream of the first half-wave plate in the optical path, for example by the M3 mirror. Polarized common aberrations downstream of the half-wave plate are removed by taking the difference of the two IRDIS channels which may provide a polarimetric precision below 0.5\% \citep{langlois2014}. Recently, pupil stabilized DPI observations have also become available.

Instrumental polarization that can not be corrected with the half-wave plate switch has to be subtracted during post-processing which typically achieved by assuming that the central star is unpolarized. This has proven to be efficient for circumstellar disks that scatter light with a high degree of polarization \citep[e.g.,][]{avenhaus2014}. For a companion on the other hand, the polarimetric signal is only of the order of 1\% or less, therefore, the polarimetric accuracy is also limited by the assumption on the unpolarized photometry of the central star which might not always be valid at the level of precision that is required. In case of sufficient field rotation, the instrumental polarization can be disentangled from the stellar polarization because the stellar polarization vector rotates with the parallactic angle causing a modulation of the instrumental polarization vector which is independent of parallactic angle \citep{perrin2015}.

\subsection{Potential targets for high-contrast infrared polarimetry}\label{sec:potential_targets}

The detectability of a companion in polarized infrared light depends strongly on the disk-integrated degree of polarization which is affected by horizontal variations in the cloud deck (see Fig.~\ref{fig:clouds}), as well as the presence of circumplanetary dust (see Fig.~\ref{fig:circumplanetary}). For a spherical planet, zonally distributed clouds may provide a detectable polarization signal (e.g., model C1) but there are also cases where the degree of polarization is close to zero (e.g., model C5). Increasing the oblateness typically increases the degree of polarization but the opposite may also occur (model C1). From a technical perspective, the contrast, angular separation, and planet brightness provide important constraints for a target selection. Although polarimetry reduces the contrast between the planet and stellar flux (in case the star is unpolarized), the brightness of a planet is a factor $\sim$100 smaller in polarized intensity than in total intensity. Here, we will discuss a few potential targets for infrared polarimetry with the results from Sect.~\ref{sec:infrared_polarization} in mind.

Beta~Pic~b \citep{lagrange2009,lagrange2010} was detected with high S/N by \citet{bonnefoy2013} at 0\ffarcs46 separation and $\Delta H = 10$~mag contrast. The planet is a prime target for infrared polarimetry for several reasons. First, $\beta$~Pic~b is spinning fast \citep{snellen2014} which will flatten the planet to $f_{\rm oblate} \simeq 0.05$ (see Eq.~\ref{eq:oblateness}) given the constraints on the mass of the planet \citep{bonnefoy2013}, and rotationally-induced zonal winds may result in horizontal cloud variations. Second, the orbit is highly inclined with respect to the sky plane \citep{wang2016} which means that, assuming an obliquity of $0\degr$, we might be observing from a favorable direction. Third, the planet is surrounded by circumplanetary material which might obscure part of the planet and scatter atmospheric photons, thereby enhancing the polarized flux. A possible challenge will be disentangling a planetary polarization signal from scattered light from the debris disk. Direct observations of $\beta$~Pic~b will be possible again in early 2019, although the transit of the planet's Hill sphere in 2017 might already give the first indications of the presence of circumplanetary material \citep{wang2016}.

The four known gas giants orbiting HR~8799 have separations in the range of 0\ffarcs37--1\ffarcs73 \citep{marois2008,marois2010} and contrasts of $\sim$10.5--12.5~mag in the $H$ and $K$ bands \citep{zurlo2016}. Several authors have shown that the spectrophotometry of the planets are best explained by atmospheric models containing patchy clouds \citep[e.g.,][]{currie2011,skemer2014} and a high-altitude haze layer of submicron-sized silicate grains \citep{bonnefoy2016}. The disadvantage of the HR~8799 system is the $\sim$$30\degr$ inclination of the planet orbits \citep{wertz2017} which means that rotationally-induced oblateness and the presence of zonal clouds will only leave a minor polarization signal if the obliquity of the planets is small although a signal due to the patchy clouds might be detectable.

HD~95086b is a 5~$M_{\rm Jup}$ gas giant that was first detected in the $L'$ band at a separation of 0\ffarcs62 from the debris disk-hosting primary \citep{rameau2013a,rameau2013b}. The mid-infrared luminosity of the planet is consistent with an L/T transition object but the $H$ -- $L'$ color is very red compared to other self-luminous planets which suggests a very dusty and low surface gravity atmosphere \citep{galicher2014}. A detailed photometric and spectroscopic characterization by \citet{derosa2016} showed that atmospheric models with high photospheric dust and a surface gravity of $\log g \lesssim 4.5$ fit best the spectral energy distribution, given the constraints on the effective temperature. A detectable polarization signal is expected for an atmosphere with an increased oblateness and high-altitude dust which makes HD~95086b, as well as other companions at the L/T transition, a feasible target. However, astrometric monitoring revealed orbital motion from which an inclination of $153\degr$ was derived by \citet{rameau2016} which might be, similar to the HR~8799 planets, a disadvantage.

A different group of potential targets are very wide orbit (>100~au) planetary mass companions. For example, HD~106906b is a young companion (13~Myr) with a predicted mass of 11~$M_{\rm Jup}$ which was detected by \citet{bailey2014} at a separation of 7\ffarcs1 (650~au), far beyond the edge-on debris disk of the system \citep{kalas2015,lagrange2016}. The favorable separation and contrast \citep[$\Delta H = 8.7$~mag;][]{bailey2014} would allow for a field-stabilized DPI observation of the companion. The very red color of HD~106906b and the radially-extended point spread function measured by the \emph{Hubble Space Telescope} hint at the presence of a disk or cloud of circumplanetary dust \citep{kalas2015} which make it an interesting target for infrared polarimetry. GSC~06214-00210b might also be a feasible target along the same line of reasoning. The companion is separated by 2\ffarcs2 from the primary and has a mass close to the deuterium-burning limit \citep[$\sim$14~$M_{\rm Jup}$;][]{ireland2011}. The red color of the companion and the detection of strong Pa$\beta$ emission are indications for the presence of a circum-substellar accretion disk \citep{ireland2011,bowler2011}. An 0.15~$M_\oplus$ upper limit on the dust mass around the companion was derived by \citet{bowler2015} from ALMA observations.

\subsection{Three-dimensional radiative transfer effects}

The importance of horizontally propagating radiation increases when the horizontal optical depth gradient is comparable to or larger than the vertical optical depth gradient, or when a horizontal temperature gradient is present in the atmosphere. For example, a locally thick deck of clouds will scatter part of the upwelling thermal flux to neighboring regions where the cloud optical depth is smaller, therefore making it easier for the radiation to escape the atmosphere. Or, horizontal changes in temperature or density will result in an asymmetry in the emitted radiation field in which case the flux from low luminosity regions will be enhanced by the neighboring high luminosity regions. In those cases, 3D radiative transfer effects might be important, in particular when horizontal scattering or temperature variations are present high in the atmosphere where the gas is optically thin (see Appendix~\ref{sec:horizontal}). For example, L/T transition dwarfs show photometric and spectroscopic modulations which have been explained by horizontal variations in cloud thickness and temperature \citep{artigau2009,apai2013}.

The intrinsic 3D grid within ARTES enables scattering calculations in arbitrary density environments without having to make any assumptions or approximations on the scattering processes. For the infrared polarization models in Sect.~\ref{sec:infrared_polarization}, we used a simplified thermal structure and cloud parametrization in order to directly control the location, optical depth, and scattering properties of the clouds. This means that there is no consistency between the temperature and density structure, as well as the sizes and opacities of the cloud particles that are considered. Combining a physical atmospheric model with the scattering radiative transfer by ARTES are planned for future work, for example to model reflected light phase curves in light of future space missions such as CHEOPS \citep{fortier2014}, TESS \citep{ricker2015}, and PLATO \citep{rauer2014}.

\subsection{Final summarizing remarks}

High-contrast infrared polarimetry is a powerful technique for the characterization of self-luminous gas giant exoplanets but technically challenging. However, significant progress is expected in this field with the installment of the latest generation of high-contrast imaging instruments. Although the integrated degree of polarization of a slowly-rotating planet requires an extreme cloud deck variation to reach 1\%, an increased oblateness may enhance the polarized flux to detectable levels depending on the horizontal distribution and thickness of the clouds. The presence of a circumplanetary disk or ring system can also introduce a significant polarization variation both through reflection of atmospheric photons, scattering of thermal disk photons, and obscuration of the planet. Although an infrared polarization detection will provide already some information about the asymmetry of a planet, a large variety of integrated scattering asymmetries is evidently possible. Therefore, the strength of high-contrast infrared polarimetry to characterize planet atmospheres will fully unravel with combined photometric and spectroscopic constraints.

\begin{acknowledgements}

We would like to thank Allona Vazan, Julien Milli, and Remco de Kok for insightful discussions. This research made use of Astropy, a community-developed core Python package for Astronomy \citep{astropy2013}.

\end{acknowledgements}

\bibliographystyle{aa}
\bibliography{references}

\appendix

\section{Benchmark results}\label{sec:benchmarks}

\begin{figure}
\resizebox{\hsize}{!}{\includegraphics{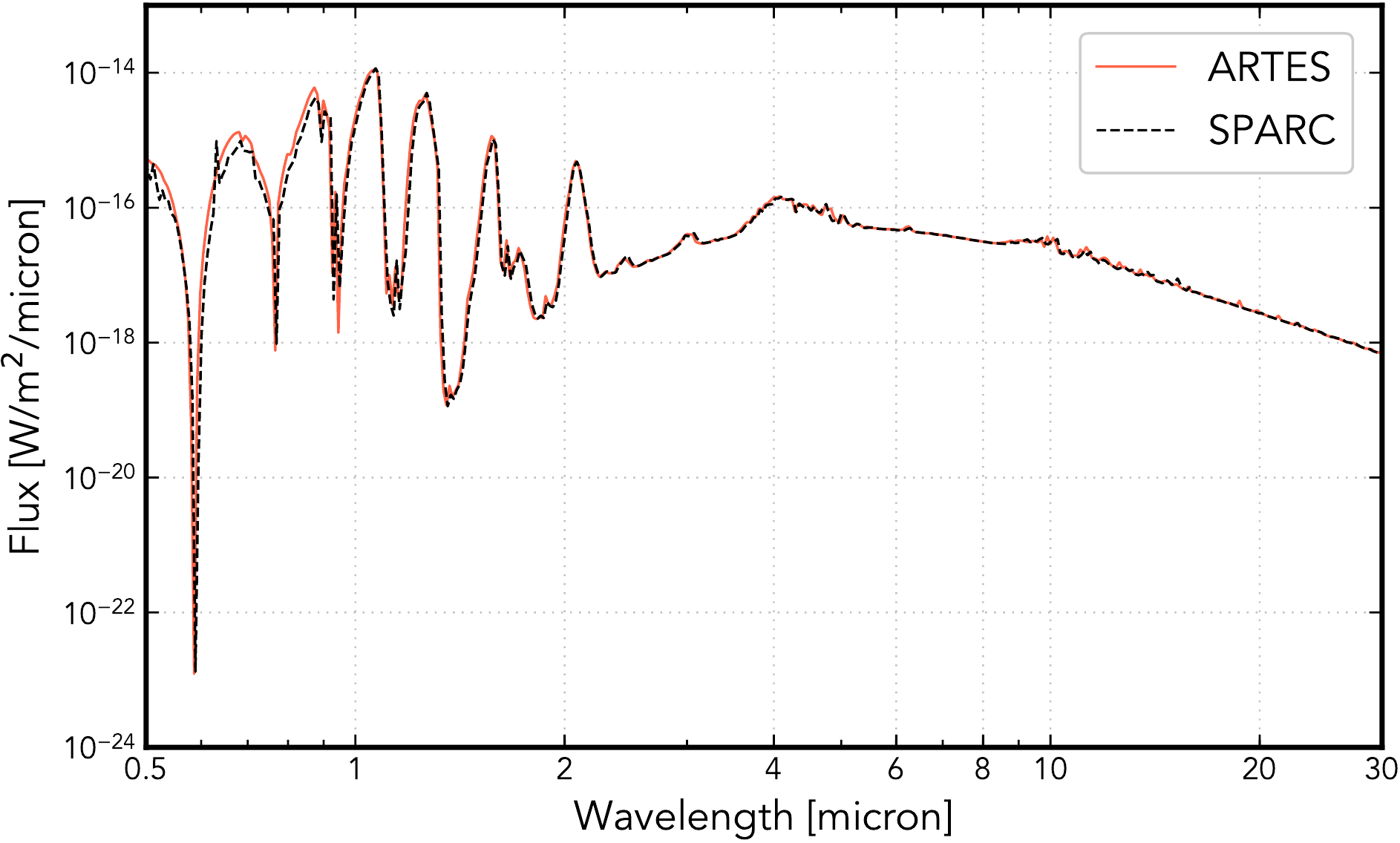}}
\caption{Emission spectrum from a self-luminous, cloudless gas giant with a gray pressure-temperature profile ($T_{\rm eff} = 700$~K). The 3D Monte Carlo radiative transfer calculation with ARTES (red solid line) is compared with the 1D atmospheric retrieval code SPARC (black dashed line). The Monte Carlo errors are smaller than the line width.\label{fig:benchmark_emission_spectrum}}
\end{figure}

In this appendix, we present the results of several benchmark calculations that are done to test the ARTES radiative transfer code. The benchmarks include an emission spectrum from a self-luminous gas giant, a reflected light spectrum from a Jupiter-like gas giant, and multiple reflected light phase curves of Rayleigh and Mie scattering atmospheres.

\subsection{Emission spectrum}\label{sec:benchmark_emission_spectrum}

As a check for the thermal radiative transfer with ARTES, we have calculated an emission spectrum from a self-luminous gas giant at 10~pc for which we have neglected the external radiation field from the star (i.e., assuming a large orbital radius). The atmosphere consists of 40 homogeneous $P$-$T$ layers (no latitudinal or longitudinal structure) with a constant mean molecular weight, $\mu=2.3$, of the gas. The $P$-$T$ profile is calculated with the gray atmosphere approximation (see Eq.~\ref{eq:gray_atmosphere}) for which we used $T_{\rm eff} = 700$~K and $\kappa=0.03$~cm$^2$/g, and a Jupiter-like surface gravity. Absorption cross sections are calculated with a spectral resolution of $\lambda/\Delta\lambda=100$ and the mixing ratios of the gaseous molecules and atoms are fixed by equilibrium chemistry (see Sect.~\ref{sec:cloudless}).

The emission spectrum is computed in the wavelength range of 0.5--30~$\mu$m, both with ARTES (3D Monte Carlo radiative transfer) and SPARC (1D atmospheric retrieval code; Min et al., in prep.). The standard setup of SPARC uses the correlated-$k$ method for the radiative transfer, but for the benchmark calculation we used averaged absorption opacities in order to be able to directly compare the model spectra of ARTES and SPARC. The spectra in Fig.~\ref{fig:benchmark_emission_spectrum} are overall in excellent agreement with slight deviations in the center of the 0.9--1.1~$\mu$m water vapor absorption band and the continuum flux in the optical. The minor offset of the continuum flux could be caused by small differences in the implementation of scattering which contributes to the emergent spectrum only at the shortest optical wavelengths where the Rayleigh scattering cross section becomes comparable to or larger than the absorption cross section of the gas. SPARC calculates the cross section with the correct weighting of the refractive indices of all molecules that are present whereas ARTES uses molecular hydrogen only.

\subsection{Reflected light spectrum}\label{sec:benchmark_reflected_spectrum}

\begin{figure}
\resizebox{\hsize}{!}{\includegraphics{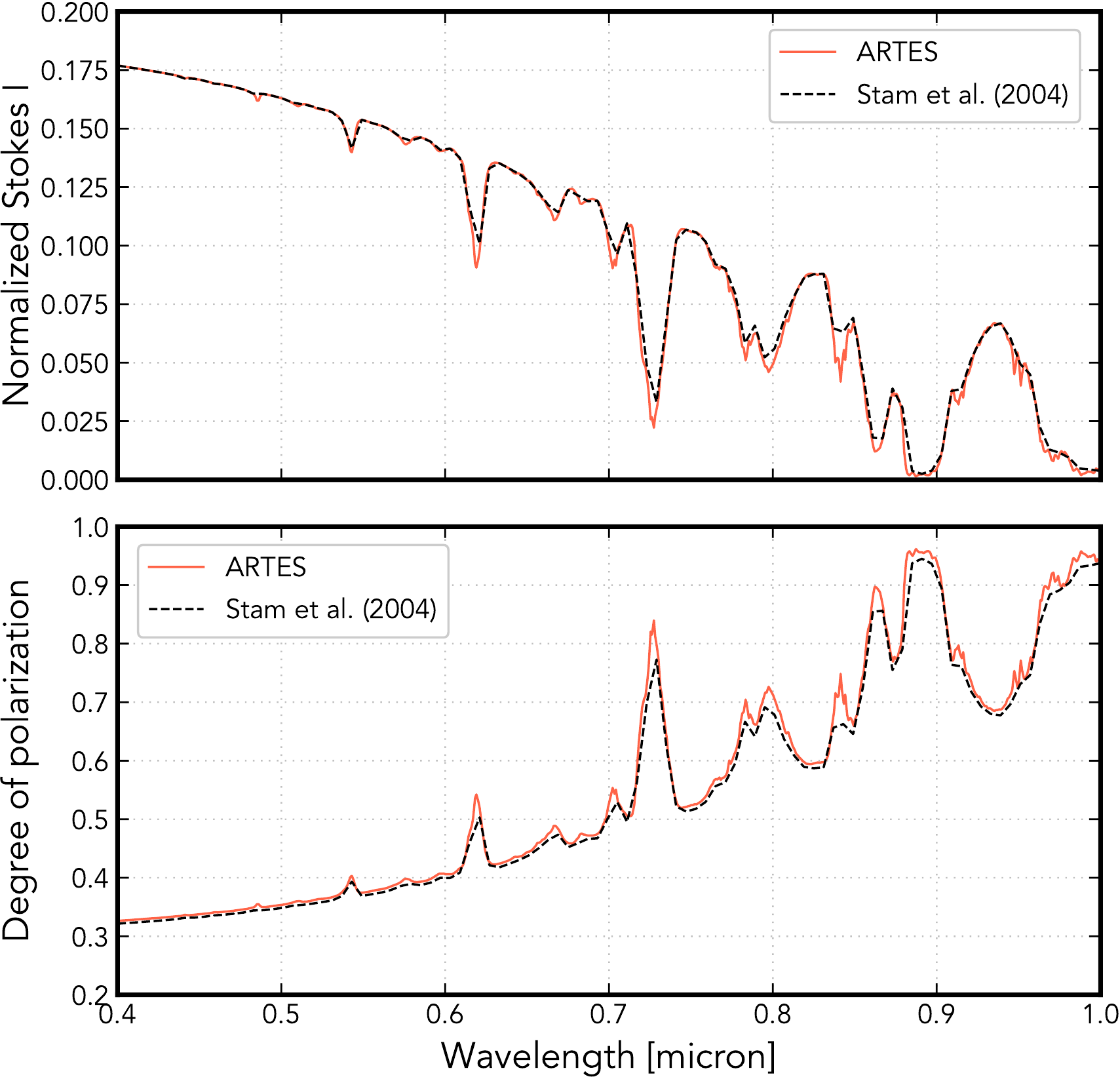}}
\caption{Reflected light spectrum of a cloudless Jupiter-like atmosphere at a phase angle of $90\degr$. The panels show the total intensity (\emph{top}) and the degree of polarization (\emph{bottom}). The ARTES spectra (red solid line) are benchmarked with the results from \citet{stam2004} (black dashed lines). The Monte Carlo errors are smaller than the line widths.\label{fig:benchmark_reflected_spectrum}}
\end{figure}

In addition to the emission spectrum, we have calculated a reflected light spectrum to further investigate the correctness of the wavelength-dependent radiative transfer calculations with ARTES. Here, we used the cloudless Jupiter-like atmosphere from \citet{stam2004} which is dominated by H$_2$ but contains CH$_4$ with a mixing ratio of $1.8\times10^{-3}$. The methane absorption coefficients are taken from \citet{karkoschka1994}, Rayleigh scattering cross sections are calculated for H$_2$, and we used a depolarization factor of 0.02 for the molecular hydrogen \citep{hansen1974b}.

In contrast to \citet{stam2004}, we did not use Jupiter's \mbox{$P$-$T$ profile} to calculate the optical thickness of the atmosphere at each wavelength, but instead we use a single layer atmosphere with a constant density. This is justified because this is a cloudless model with constant opacities throughout the atmosphere. Therefore, only the total vertical optical depth affects the reflected light spectrum and not the vertical density structure. We scaled the density to match the 21.47 optical depth from \citet{stam2004} at 0.4~$\mu$m after which the same value of the density is used for all other wavelengths.

The surface albedo at the inner boundary of the grid is set to $A=0.0$ such that all crossing photons are absorbed. The detector is located at a phase angle of $\alpha=90\degr$, that is, the phase angle for which the single scattering polarization of Rayleigh scattering is maximal. Figure~\ref{fig:benchmark_reflected_spectrum} shows the normalized Stokes~$I$ reflected light spectrum in the wavelength range of 0.4--1.0~$\mu$m, as well as the wavelength-dependent degree of polarization, $P=-Q/I$. The spectra are compared with the results from \citet{stam2004} who use a locally plane-parallel atmospheric model with the adding-doubling technique to calculate the radiative transfer. The spectra are in good agreement although the ARTES degree of polarization spectrum shows a minor offset, consistently for all wavelengths, of which the origin is unknown. The flux is normalized by the incoming stellar flux at the substellar point of the atmosphere,
\begin{equation}\label{eq:flux_norm}
C_{\rm norm} = \pi B \frac{R_*^2 R_{\rm pl}^2}{D^2 d^2},
\end{equation}
where $\pi B$ is the Planck flux at the stellar surface, $R_*$ the stellar radius, $R_{\rm pl}$ the planet radius, $D$ the distance between the star and the planet, and $d$ the distance between the planet and the observer. In this way, the normalized flux values have no dependence on any of the parameters in Eq.~\ref{eq:flux_norm} and the value of Stokes~$I$ at a $0\degr$ phase angle corresponds with the geometric albedo \citep{stam2004}, that is, the planet brightness at $0\degr$ phase angle normalized to a fully reflecting, diffusively scattering disk of the same radius, $R_{\rm pl}$.

\subsection{Reflected light phase curves}\label{sec:benchmark_reflected_phase}

Exoplanet phase curves show the intensity of an exoplanet as function of its orbit. As a benchmark for the phase angle dependence, we calculated reflected light phase curves, both of the total intensity and polarized intensity, for atmospheres with Rayleigh scattering particles, Lambertian surface reflection, and NH$_3$ cloud particles. The ARTES phase curves are compared with results from the literature. The emergent flux from a Lambertian surface is given by
\begin{equation}\label{eq:lambert}
j(\alpha) = \frac{2}{3}\omega\pi F_*\left[ \frac{\sin{\alpha}+(\pi-\alpha)\cos{\alpha}}{\pi}  \right],
\end{equation}
where $\omega$ is the single scattering albedo, $\alpha$ the phase angle, and $\pi F_*$ the incident stellar flux. Each benchmark model consists of a single homogeneous atmospheric layer with a radial optical depth $\tau$, surface albedo $A$, and single scattering albedo $\omega$.

\begin{figure}
\resizebox{\hsize}{!}{\includegraphics{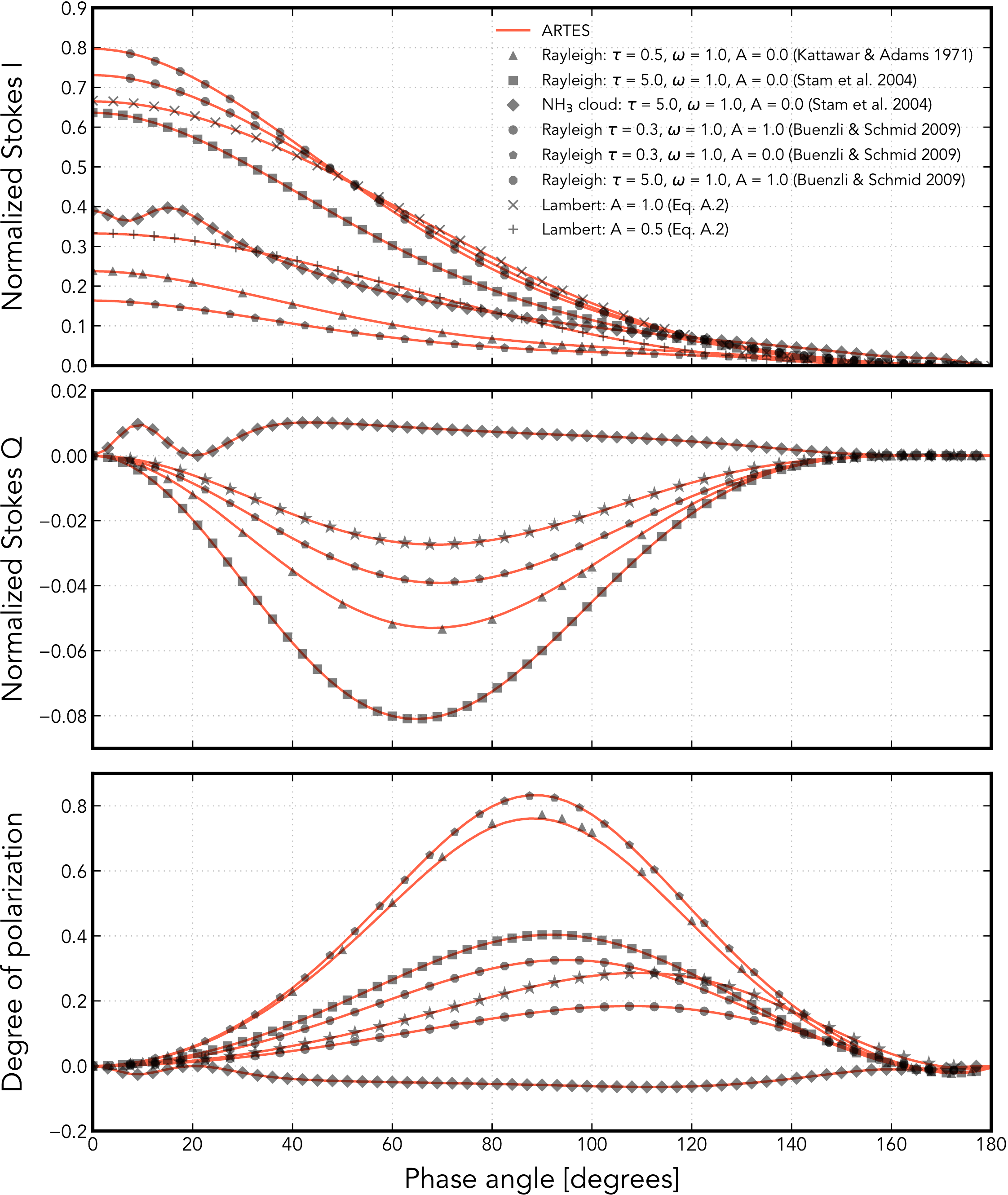}}
\caption{Reflected light phase curves of atmospheres with Rayleigh or Mie scattering particles for various optical depths, surface albedos, and single scattering albedos. The panels show the normalized Stokes~$I$ (\emph{top}), normalized Stokes~$Q$ (\emph{center}), and degree of polarization (\emph{bottom}). The phase curves computed with ARTES (red solid lines) are benchmarked with results from \citet{buenzli2009}, \citet{stam2004}, \citet{kattawar1971a}, and the Lambertian surface reflection (symbols). The Monte Carlo errors are smaller than the line widths.\label{fig:benchmark_reflected_phase}}
\end{figure}

The scattering properties of the ammonia ice particles are calculated with Mie theory \citep{min2005,toon1981} at a wavelength of 0.7~$\mu$m and the real and imaginary part of the complex refractive index are set to $n=1.42$ and $k=10^{-6}$, respectively \citep{martonchik1984}. Therefore, the single scattering albedo of the ammonia cloud particles is approximately unity. For the size distribution, we used an effective radius and variance of 1.0~$\mu$m and 0.1, respectively (see Eq.~\ref{eq:gamma_distribution}). The opacities and scattering matrices are calculated with two different codes which gave identical results \citep{derooij1984,min2005}.

Figure~\ref{fig:benchmark_reflected_phase} shows the benchmark results of the normalized Stokes~$I$, normalized Stokes~$Q$, and degree of polarization phase curves for various optical depths, surface albedos and single scattering albedos. The Rayleigh and Mie scattering calculations with ARTES are compared with the locally plane-parallel, doubling-adding calculations by \citet{stam2004}, as well as the Monte Carlo radiative transfer calculations by \citet{buenzli2009} and \citet{kattawar1971a}. The Lambertian surface reflection calculations are compared with the analytical solution from Eq.~\ref{eq:lambert}. All phase curves are in good agreement with each other. As expected, increasing the optical depth or surface albedo results in a larger Stokes~$I$ but smaller Stokes~$Q$ because multiple scattering dampens the degree of polarization. We note that the planet integrated Stokes~$U$ is zero because the model atmospheres are symmetric with respect to the single scattering plane. We follow \citet{stam2004} with their definition for the degree of polarization, $P=-Q/I$, such that a positive value corresponds with light which is polarized perpendicular to the single scattering plane.

\section{Atmospheric model spectra}\label{sec:model_spectra}

\begin{figure}
\resizebox{\hsize}{!}{\includegraphics{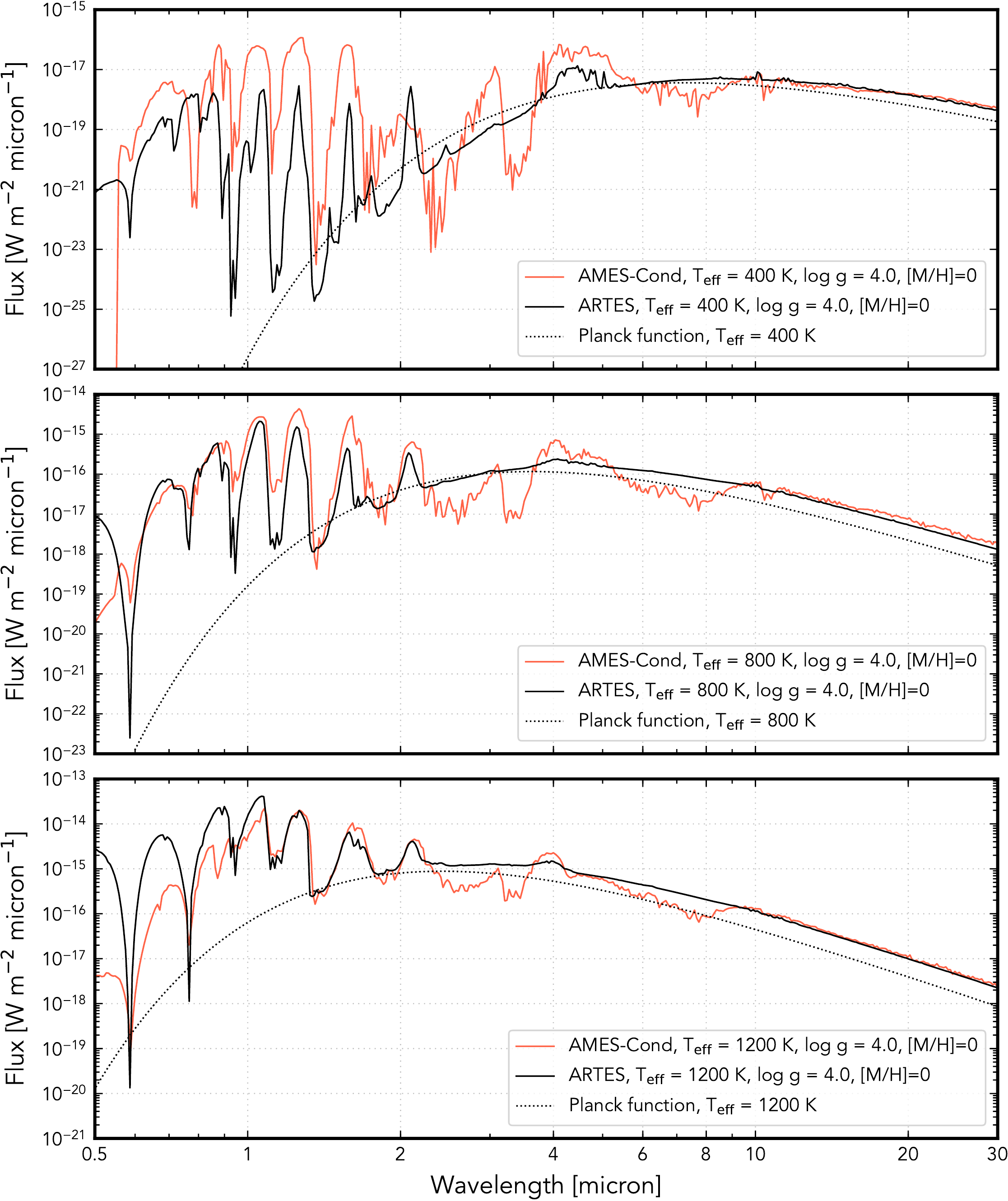}}
\caption{Comparison of the calculated emission spectra from Fig.~\ref{fig:self_luminous} with the AMES-Cond atmospheric models by \citet{allard2001}, both shown with a spectral resolution of $\lambda/\Delta\lambda=100$. The panels display, from top to bottom, the emission spectra for $T_{\rm eff} = 400$~K, $T_{\rm eff} = 800$~K, and $T_{\rm eff} = 1200$~K. Fluxes have been scaled to a planet with a radius of 1.3~$R_{\rm Jup}$ at a distance of 10~pc.\label{fig:model_spectra}}
\end{figure}

In this appendix, we provide a comparison of the calculated emission spectra with those from the AMES-Cond atmospheric models by \citet{allard2001} which include detailed physics and chemistry. Dust condensation occurs in the AMES-Cond models in chemical-equilibrium with the gas, but the effect of the dust opacities has been neglected. Therefore, the models correspond to an atmosphere in which dust grains have settled below the photosphere and allow for an approximate comparison with the cloudless atmosphere models from Sect.~\ref{sec:cloudless}.

Figure~\ref{fig:model_spectra} displays the emission spectra of the AMES-Cond models, binned to a spectral resolution of $\lambda/\Delta\lambda=100$, in comparison with the ARTES spectra from Fig.~\ref{fig:self_luminous}. The effective temperature, \mbox{$T_{\rm eff}=\{400,800,1200\}$~K}, surface gravity, $\log{g} = 4.0$, and metallicity, ${\rm [M/H]} = 0.0$, are set to identical values. Differences are to be expected as the thermal structure, gas opacities, and mixing ratios are self-consistently calculated in the AMES-Cond models, while an gray atmosphere approximation is used for the thermal structure of the ARTES models. For example, the ARTES spectra appear featureless in the wavelength range of \mbox{$\sim$2.1--4.0~$\mu$m} as a result of the highly isothermal upper part of the atmosphere (see $P$-$T$ profiles in Fig.~\ref{fig:self_luminous}). Also, different line lists are used to calculate the gas opacities which will have an effect on the emission spectra.

\section{Horizontal radiation transport}\label{sec:horizontal}

Scattering radiative transfer calculations in planetary atmospheres are typically done with locally plane-parallel atmospheric models \citep{stam2004,stam2008,spurr2006,buenzli2009}. Horizontal inhomogeneities are incorporated by dividing the signal from the planetary disk into a collection of independent pixels on the planet \citep[e.g.,][]{dekok2011,karalidi2012,karalidi2013} because typically the differential transport of horizontally propagating radiation is negligible. However, there are scenarios in which horizontal radiation transport might affect the planet luminosity.

One scenario is when the optical depth gradient is steeper in horizontal direction than it is in vertical direction. A fraction of the radiation will propagate preferentially in horizontal direction toward the low optical depth region in order to escape from the atmosphere. This may for example occur when the vertically upward energy flow is locally hindered by the presence of thick clouds. Another scenario is when horizontal temperature variations are present that cause an asymmetry in the radiation field, for example due to a hot spot or in the proximity of the day-night terminator of a tidally locked planet. In that case, the flux from the low luminosity regions will be enhanced by the flux coming from the high luminosity regions.

\begin{figure}
\resizebox{\hsize}{!}{\includegraphics{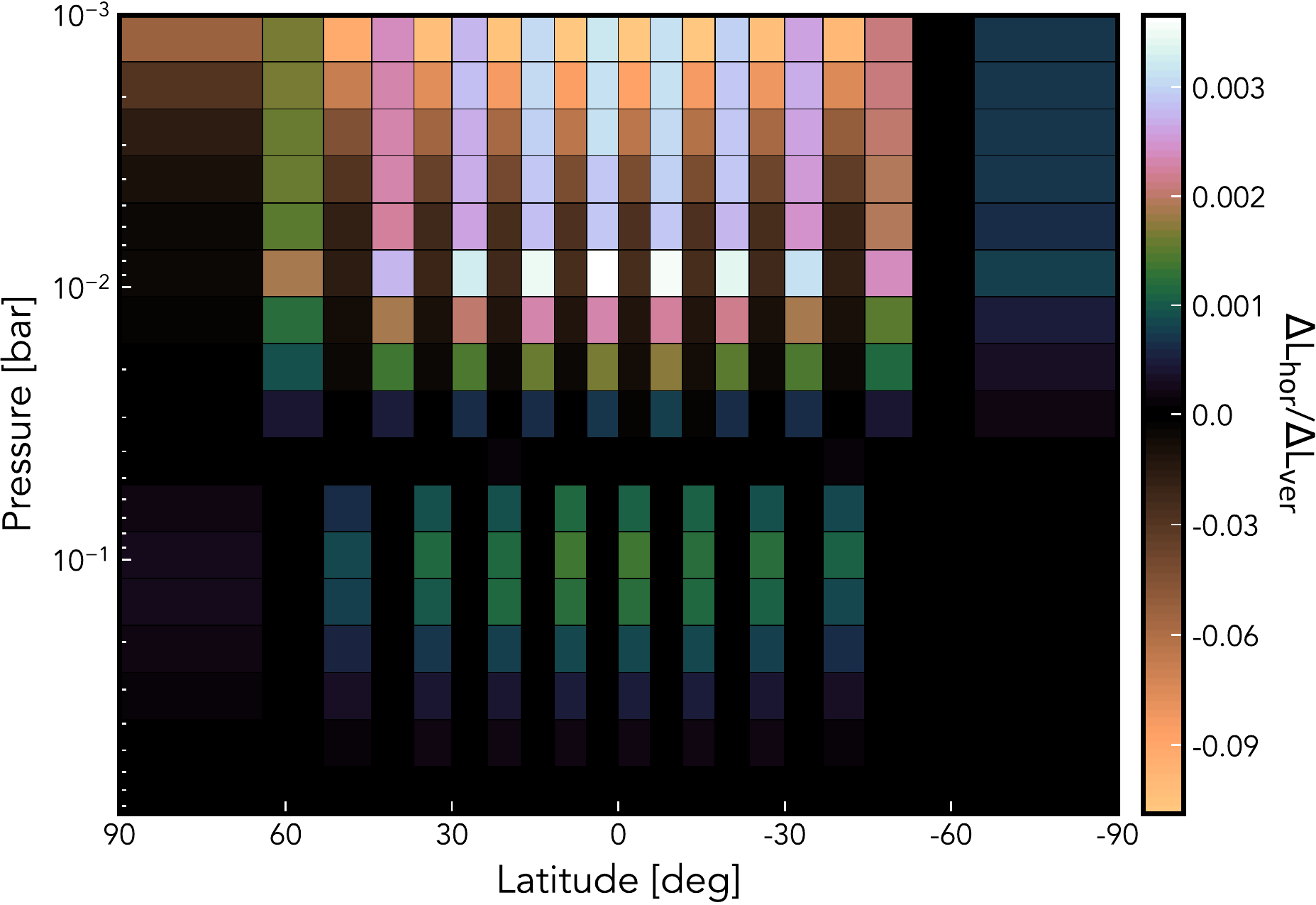}}
\caption{Net horizontal radiation transport in an atmosphere with high-altitude (10~mbar) zonal clouds, normalized to the net vertical energy transport. The negative part of the color bar has a larger dynamical range than the positive part.\label{fig:horizontal}}
\end{figure}

The effect of scattering by non-uniform clouds on the differential radiation transport is investigated with ARTES. We use a gray $P$-$T$ profile ($T_{\rm eff}=800$~K, $\log{g}=3.4$) and latitudinal variations in the distribution of clouds. As a proof of concept, we parameterized the latitudinal cloud distribution with 20~linearly spaced values of $\sin{\theta_{\rm lat}}$ (with $\theta_{\rm lat}=[-90,90]$ the latitude), with alternating optical depth ($\tau_1=0$ and $\tau_2=5$) and located at high altitude ($P_{\rm cloud} = 10$~mbar). We use an isotropic phase function, $P(\cos{\Theta}) = 1/2$, for the cloud particles, and all other elements of the scattering matrix are set to zero. The horizontal radiation transport is measured by keeping track of all the energy that is propagating through the latitudinal cell boundaries of the atmospheric grid in northern and southern direction.

Figure~\ref{fig:horizontal} displays the net horizontal transport (i.e., difference between energy flowing horizontally out and into a grid cell) as function of altitude and latitude in the atmosphere, normalized to the net vertical transport through each grid cell. The differential flow is positive at the latitudes with clouds, and it transitions to zero around 50~mbar where the atmosphere becomes optically thick. At latitudes with no clouds, the net horizontal flow is negative, that is, a larger fraction of energy is horizontally entering the cells instead of leaving. Part of the photons are scattered downwards which results in a positive flow in the deeper atmospheric regions.

The differential transport will reduce when clouds are located at lower altitudes where the optical depth from the gas is higher. In that case, the horizontal optical depth increases with respect to the vertical optical depth and most photons can escape more easily in upward direction than by crossing a latitudinal cell boundary. Only horizontal variations in the scattering optical depth are considered in this example whereas horizontal temperature gradients will enhance the effect. \citet{showman2013} determined with a 3D global circulation model of a brown dwarf atmosphere that hydrodynamically-induced horizontal temperature variations can be as large as $\Delta T = 50$~K which will locally cause fractional flux variations of $\Delta F/F \sim $~0.02--0.2.

\end{document}